\documentclass{aa}
\usepackage{graphicx}
\usepackage{txfonts}
\usepackage{color}
\newcommand{\cmtwo}{cm$^{-2}$}  
\newcommand{\cmthree}{cm$^{-3}$}

\newcommand{\kms}{km\,s$^{-1}$}       

\newcommand{\vlsr}{$\upsilon_{\rm LSR}$}        

\newcommand{\ta}{$T_{\rm A}$}        

\newcommand{\tas}{$T^{*}_{\rm A}$}

\newcommand{\tmbdv}{$\int\!T_{\rm mb}\,d\upsilon$}
\newcommand{\tmb}{$T_{\rm mb}$}
\newcommand{\trms}{$T_{\rm rms}$}
\newcommand{\tsys}{$T_{\rm sys}$}
   
\newcommand{\tdv}{$\int\!T\,d\upsilon$}

\newcommand{\ecs}{erg cm$^{-2}$ s$^{-1}$}

\newcommand{\um}{$\mu$m}                                 
\newcommand{\molh}{H$_{2}$}                              

\newcommand{\water}{H$_{2}$O}
\newcommand{\etttio}{{\rm H}$_2$O$\,(1_{10} - 1_{01})$}

\newcommand{\molo}{O$_{2}$}                     

\newcommand{\coiso}{${\rm C}^{18}{\rm O}$}

\newcommand{\dcop}{DCO$^{+}$}
\newcommand{\msun}{$M_{\odot}$}

\newcommand{\gapprox}{$\stackrel {>}{_{\sim}}$}   
\newcommand{\lapprox}{$\stackrel {<}{_{\sim}}$}
\newcommand{\about}{$\sim$}                       
\newcommand{\powten}[1]{10$^{#1}$}
\newcommand{\oishort}{[O\,{\sc i}]\,63\,$\mu$m}
\newcommand{\oilong}{[O\,{\sc i}]\,145\,$\mu$m}

\newcommand{\pzero}{$^{3}{\rm P}_{0}$}
\newcommand{\pone}{$^{3}{\rm P}_{1}$} 
\newcommand{\ptwo}{$^{3}{\rm P}_{2}$}

\newcommand{\ammonia}{{\rm NH}$_3$}     

\newcommand{\ettnoll}{{\rm NH}$_3\,(1_0 - 0_0)$}
\newcommand{\ntwohp}{${\rm N_2H^+}$}
\newcommand{\av}{$A_{\rm V}$}                     
\newcommand{\rv}{$R_{\rm V}$}

\newcommand{\ro}{$\rho \, {\rm Oph}$}
\newcommand{\roc}{$\rho \, {\rm Oph \,\, cloud}$}
\newcommand{\roa}{$\rho \, {\rm Oph \, A}$}

\newcommand{\amin}{$^{\prime}$}                   
\newcommand{\asec}{$^{\prime \prime}$}
\newcommand{\adeg}{$^{\circ}$}

\newcommand{\radot}[4]{\mbox{#1$^{\rm h}$#2$^{\rm m}$#3$\stackrel {\rm s}{_{\bf\cdot}}$#4}}  
\newcommand{\decdms}[3]{\mbox{#1$^{\circ}$#2$^{\prime}$#3$^{\prime \prime}$}}

\newcommand{\adegdot}[2]{\mbox{#1$\stackrel {\circ}{_{\bf \cdot}}$#2}}
\newcommand{\amindot}[2]{\mbox{#1$\stackrel {\prime}{_{\bf \cdot}}$#2}}
\newcommand{\asecdot}[2]{\mbox{#1$\stackrel {\prime \prime}{_{\bf \cdot}}$#2}}

\begin{document}

\title{Gas and dust in the star-forming region $\rho$ Oph A
\thanks{Based on observations with {\it Herschel} which is an ESA space observatory with science instruments provided by European-led Principal Investigator consortia and with important participation from NASA.}$^{,}$
\thanks{The data cubes of Figures 3 to 10 and 12 and A.1 are only available in electronic form at the CDS via anonymous ftp to cdsarc.u-strasbg.fr (130.79.128.5) or via http://cdsweb.u-strasbg.fr/cgi-bin/qcat?J/A+A/}
}

\subtitle{II. The gas in the PDR and in the dense cores}

\author{
                B. Larsson\inst{1}                                              
        \and
                R. Liseau \inst{2} 
        }

  \institute{   Department of Astronomy, Stockholm University, SE-106 91 Stockholm, Sweden,  \email{\small{bem@astro.su.se}} 
        \and 
                Department of Space, Earth and Environment, Chalmers University of Technology, Onsala Space Observatory, SE-439 92 Onsala, Sweden,  \email{\small{rene.liseau@chalmers.se}} 
        }

\date{Received ... / Accepted ...}
%
%

\abstract
{The evolution of interstellar clouds of gas and dust establishes the prerequisites for star formation. The pathway to the formation of stars can be studied in regions that  have formed stars, but which at the same time also display the earliest phases of stellar evolution, i.e. pre-collapse/collapsing cores (Class -1), protostars (Class 0), and young stellar objects (Class I, II, III).}
{We  investigate to what degree local physical and chemical conditions are related to the evolutionary status of various objects in star-forming media.}
{\roa\ displays the entire sequence of low-mass star formation in a small volume of space. Using spectrophotometric line maps of \molh, \water, \ammonia, \ntwohp, O$_2$, O\,I, CO, and CS, we examine the distribution of the atomic and molecular gas in this dense molecular core. 
The physical parameters of these species are derived, as are their relative abundances in \roa. 
Using radiative transfer models, we examine the infall status of the cold dense cores from their resolved line profiles of the ground state lines of \water\ and \ammonia, where for the latter no contamination from the VLA 1623 outflow is observed and line overlap of the hyperfine components is explicitly taken into account.}
{The stratified structure of this photon dominated region (PDR), seen edge-on, is clearly displayed. Polycyclic aromatic hydrocarbons (PAHs) and O\,I are seen throughout the region around the exciting star S\,1. At the interface to the molecular core 0.05\,pc away, atomic hydrogen is rapidly converted into \molh, whereas O\,I protrudes further into the molecular core. This provides oxygen atoms for the gas-phase formation of \molo\ in the core SM\,1, where $X$(\molo)$\,\sim 5 \times 10^{-8}$. There, the ratio of the \molo\ to \water\ abundance [$X$(\water)$\,\sim 5 \times 10^{-9}$] is significantly higher than unity. Away from the core, \molo\ experiences a dramatic decrease due to increasing \water\ formation. Outside the molecular core \roa, on the far side as seen from S\,1, the intense radiation from the 0.5\,pc distant early B-type star HD\,147889 destroys the molecules. }
{Towards  the dark core SM\,1, the observed abundance ratio $X$(\molo)/$X$(\water)$\,>1$, which suggests that this object is extremely young, which would explain why \molo\ is such an elusive molecule outside the solar system.}

\maketitle
%

\begin{figure*}
  \resizebox{\hsize}{!}{
  \rotatebox{90}{\includegraphics{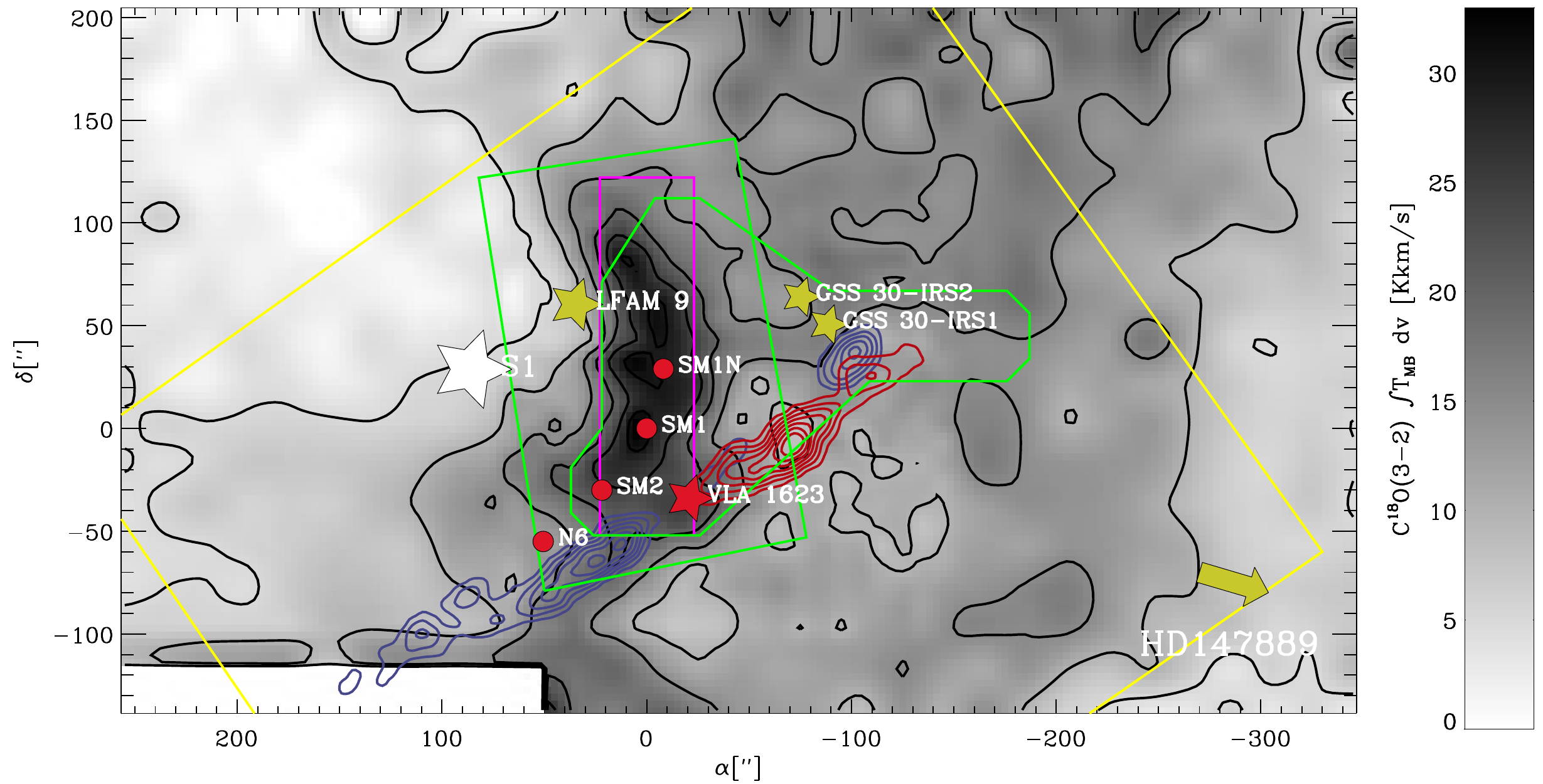}}
  }
   \caption{ Regions of our {\it Herschel} observations in \roa\ are superposed onto a greyscale and contour map of \coiso\,(3-2) integrated intensity; see scale bar to the right \citep{liseau2010}. The region mapped with HIFI in \water\,557\,GHz is shown by the large and partially cut rectangle (yellow; see also Fig.\,\ref{odin_map}) and that in \water\,1670\,GHz by the small rectangle (magenta). The green rectangle outlines the region mapped with PACS in the [O\,I]\,63,\,145\,\um\ lines and in the far-IR continuum. The green polygon marks the perimeter of the \molo\,487\,GHz observations. Yellow stars identify Class\,I and II sources, the red star the Class\,0 source VLA\,1623, and the white star the early-type object S\,1 \citep[B4 + K,][]{gagne2004}. An arrow points towards the 11\amin\ distant early B star HD\,147889. The red and blue contours refer to high-velocity gas seen in CO\,(3-2) emission (JCMT archive). Zero offset is at the position of SM\,1, i.e. for J\,2000 coordinates R.A.\,=\,\radot{16}{26}{27}{9}, Dec.\,=\,\decdms{$-24$}{23}{57}.}
   \label{obs_outlay}
\end{figure*}

\begin{figure*}
  \resizebox{\hsize}{!}{
  \rotatebox{00}{\includegraphics{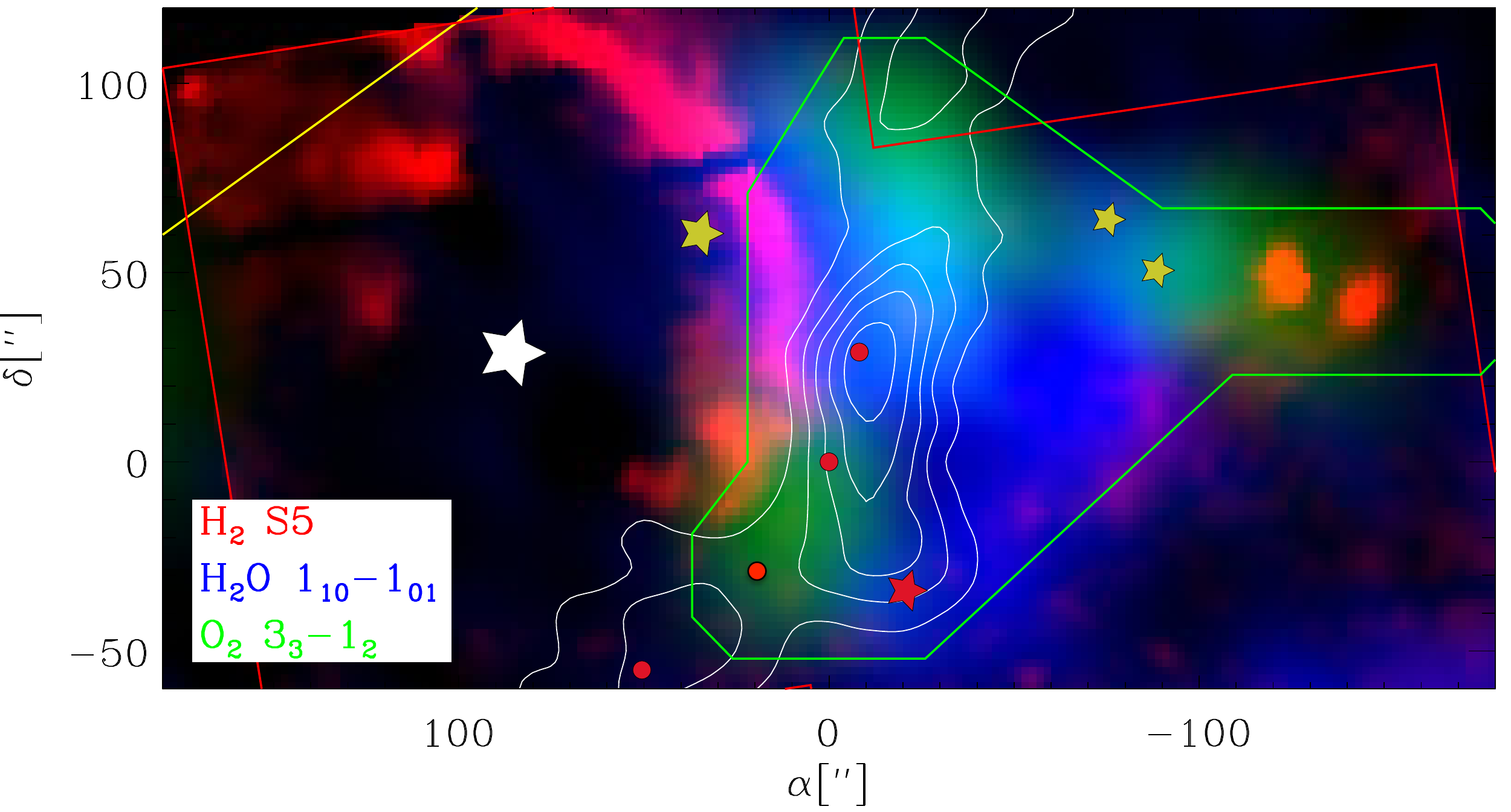}}
  }
 \caption{Observed layers of the edge-on PDR in \roa. From left to right: \molh\,6.9\,\um, \molo\,487\,GHz, and \water\,557\,GHz are shown in red, green, and blue, respectively. Symbols are as in Fig.\,\ref{obs_outlay} and the region of \molh\ observations is shown by the partial red  polygon. The \molh\ sources near (-100, 50) are gas shocked by HH-flows \citep[see also][]{liseau2009}, and faint \molh\ emission can also be seen from the VLA\,1623 outflow. The white contours  outline the integrated \ntwohp\,(3-2) emission, indicating the presence of dense gas above $10^6$\,\cmthree\  \citep[][Paper\,I]{liseau2015}.
        }
  \label{pdr}
\end{figure*}

\begin{table*}
\caption{Instrumental reference: {\it Herschel}}            
\label{Herschel}      
\begin{tabular}{llcccc}    
\hline\hline    
\noalign{\smallskip}             
Species and                     & Frequency                             & Wavelength              & Energy                        & Beam width            & Main beam                 \\
Transition                              &  $\nu_0$ (GHz)                        & $\lambda_0$ (\um)& $E_{\rm up}/k$ (K)   & HPBW (\asec)  & efficiency $\eta_{\rm mb}$     \\
\noalign{\smallskip}    
\hline                          
\noalign{\smallskip}    
{\bf PACS}                      &                                               &                               &                               &                               &       \it{a}   \\
\noalign{\smallskip}    
\hline                          
[O\,I]\,(\pone\ - \ptwo)         & 4744.738586                  & \phantom{1}63 &  228                    &\phantom{1}7           & 0.73   \\
\hspace*{7.5mm}(\pzero\ - \pone) &      2060.052199             & 145                   & 326                     & 12                            & 0.54   \\
\water\,($2_{12}-1_{01}$) &1669.904775                          & 179                   & 114                     & 13                            & 0.30   \\
\noalign{\smallskip}
\hline
\noalign{\smallskip} 
{\bf HIFI}                              &                                               &                               &                               &                               &  \it{b}      \\
\noalign{\smallskip}    
\hline                          
\water\,($2_{12}-1_{01}$)       &1669.904775                            & 179                     & 114                   & 13                            & 0.55    \\
C$^{17}$O\,($7-6$)              & \phantom{1}786.2808166        & 381                   & 151                     & 26                            & 0.60  \\
\molo\,($5_4-3_4$)              &  \phantom{1}773.839512        & 388                   &  \phantom{1}61  & 27                            & 0.60  \\
$^{13}$CO\,($7-6$)              & \phantom{1}771.184125          & 389                  & 148                     & 27                            & 0.60  \\
\ettnoll                                &  \phantom{1}572.498160        & 524                     &  \phantom{1}27        & 36                            & 0.62     \\
\ntwohp\,($6-5$)                &  \phantom{1}558.957500        & 536                   &  \phantom{1}94  & 37                            & 0.62  \\      
\water\,($1_{10}-1_{01}$) &  \phantom{1}556.9359877     & 538                   &  \phantom{1}61  & 37                            & 0.62           \\
CS\,($10-9$)                    & \phantom{1}489.7509210        & 612                   & 129                     & 42                            & 0.62  \\
\molo\,($3_3-1_2$)              & \phantom{1}487.249264         & 616                   & \phantom{1}26   & 44                            & 0.62  \\
\noalign{\smallskip}
\hline                                   
\end{tabular}
        \begin{list}{}{}
    \item[$^{a}$] PCalSpectrometer\_Beam\_v6: for 62, 45 and 168/187\,\um\ spatial averages are shown.
    \item[$^{b}$] Mueller et al.\,(2014), Release Note \#1, HIFI-ICC-RP-2014-001, v1.1.
    \end{list}
\end{table*}

%

\section{Introduction}

The physics and chemistry of star-forming clouds are governed by the competition between the energy input into the cloud and the energy lost through the radiation from the gas and the dust. This ability to cool efficiently is paramount during the gravitational formation of stars and planets and, on a grander scale, of galaxies. The abundance of the major coolants is determined by feedback processes between the atoms, molecules, and dust grains that are present in the media. However, many of these processes are only partially understood and are the focus of intense contemporary research. In this paper, we  attempt to disentangle the various interrelations among key atomic and molecular species by analysing their distribution within a star-forming cloud. 

We chose as a prime example the nearest region of star formation, i.e. the molecular cloud L\,1688 at a  distance of 120\,pc \citep{loinard2008}. Major properties of the dark clouds in the constellation of Ophiuchus have been reviewed by \citet{wilking2008} and, more recently, by \citet{white2015}.  A dense stellar cluster has been forming over the past one to two million years \citep{bontemps2001,evans2009,ducourant2017}, transforming cloud matter into a young stellar population at an overall efficiency of a few percent, but with local excursions to a few tens of percent \citep{wilkinglada1983,liseau1995,liseau1999,bontemps2001,evans2009}. Several intensity maxima in \dcop\ line emission were named `cores' and assigned the capital letters A through F by \citet{loren1990}. Since then, this scheme has been extended  \citep[e.g.][]{white2015,punanova2016}.

The dense core \roa\ attracts our particular attention as it harbours several manifestations of star formation within a relatively small volume (length scale approximately 0.1\,pc), namely a number of pre-stellar cores, a Class\,0 source with its highly collimated molecular jet, and Class\,I/II sources with their outflows and HH-objects \citep[Fig.\,\ref{obs_outlay} and][]{liseau2015}. In addition, \roa\ is prominent for its unique chemistry, including that of \molo, H$_2$O$_2$, and HO$_2$. These were first detected outside the solar system by our group \citep{larsson2007,liseau2012, bergman2011b,parise2012}.

We have studied \roa\ from space and the ground in a variety of spectral lines and in the continuum. The first results, focussing on the dust properties and the gas-to-dust relationship, are presented in Paper\,I  of this series \citep{liseau2015}. The subregions of \roa\ that were observed with various instruments aboard {\it Herschel} \citep{pilbratt2010} are outlined in Figure\,\ref{obs_outlay}, superposed onto an image in the C$^{18}$O\,(3-2) line;  in this paper, we investigate further the complex physical and chemical interplay between the gas and the dust in \roa, and what their relation is with regard to star formation per se.

In Sect.\,2, the observations and the reduction of the data are described. In Sect.\,3, the spatially resolved low-($G_0/n$) photon dominated region (PDR) is introduced. In Sect.\,4 these data are compared with theoretical PDR models, and the  \molo\ formation in \roa\ and possible scenarios of observed protostellar mass infall are discussed. In Sect.\,5 our main conclusions of this paper are summarized, and in Sect.\,6 a synergy of the results and conclusions of  Papers I and II are put into the wider context of star formation. Finally, our Odin mapping observations in the \etttio\ line of \roa, B1, B2, C, D, E, and F are presented in Appendix\,A.

%
%
\begin{figure*}
 \resizebox{\hsize}{!}{
  \rotatebox{00}{\includegraphics{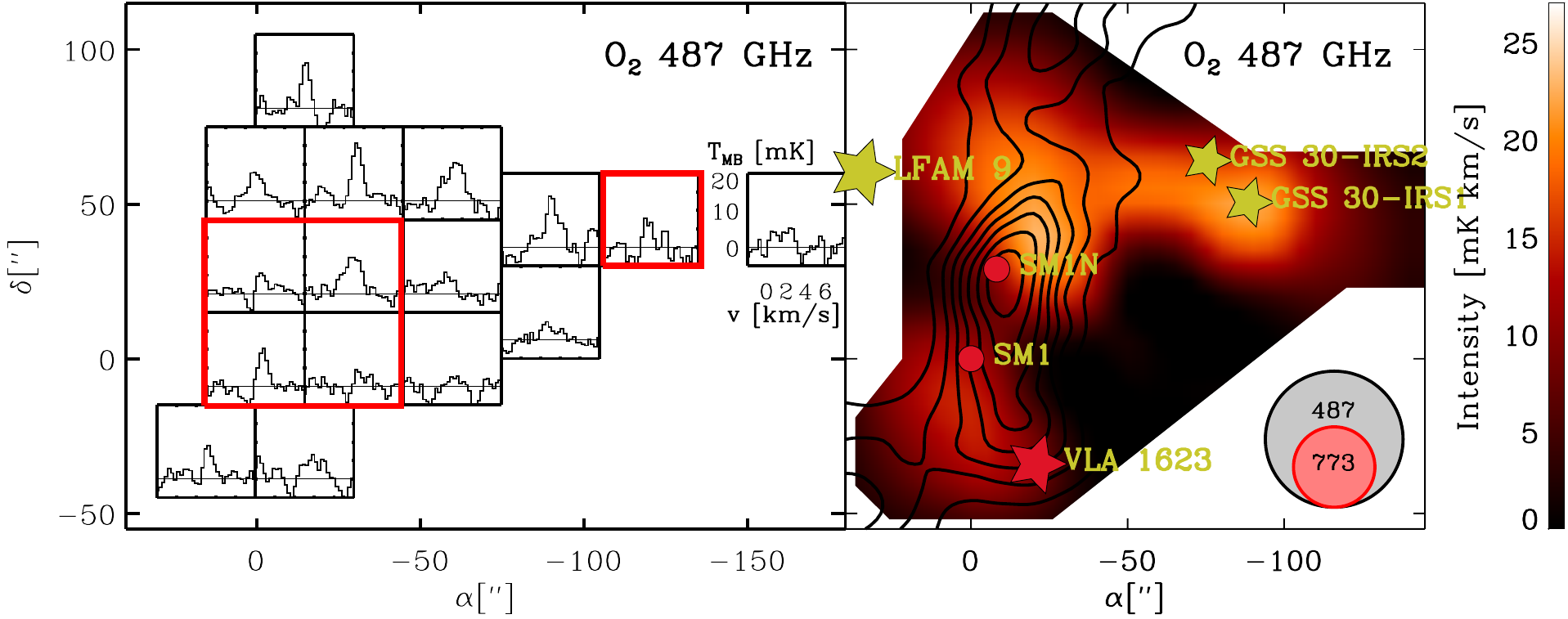}}
  }
  \caption{{\bf Left:} Map of \roa\  in the \molo\,($3_3-1_2$)\,487\,GHz line with HIFI. At the positions of the red frames oversampled \molo\,($5_4-3_4$)\,773\,GHz spectra were obtained. Scales in \kms\ and mK are indicated in the spectrum of the off-position (westernmost frame). {\bf Right:}  Contour-colour map of the data shown in the left panel. The colour bar for the integrated intensity in mK\,\kms\ is shown to the right. The crosses identify the observed positions and the contours outline integrated \ntwohp\,(3-2) emission. 
        }
  \label{o2_map}
\end{figure*}

\begin{figure*}
 \centering 
  \resizebox{\hsize}{!}{
  \rotatebox{360}{\includegraphics{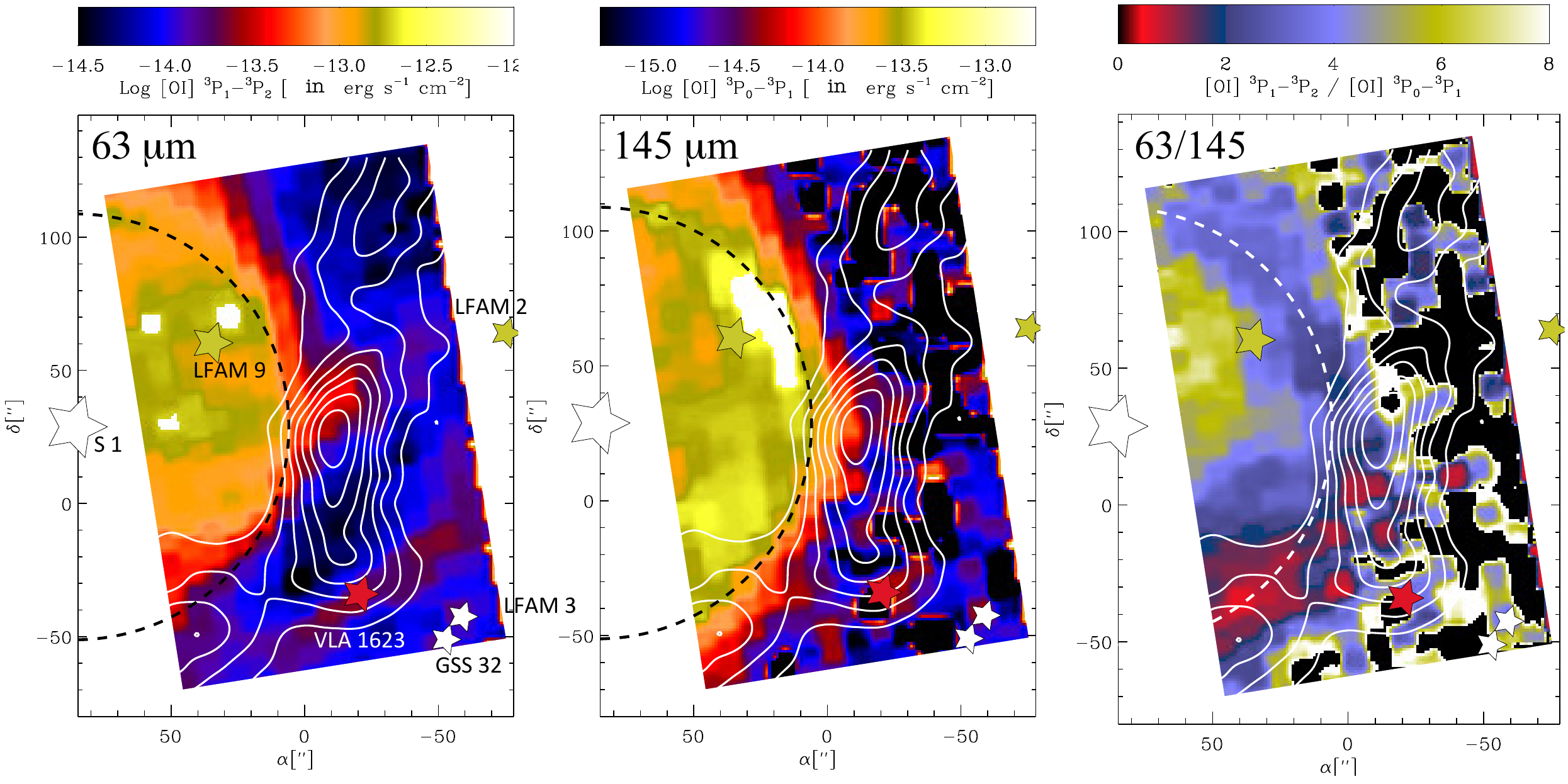}} 
  }
   \label{oi}
  \includegraphics[width=18cm]{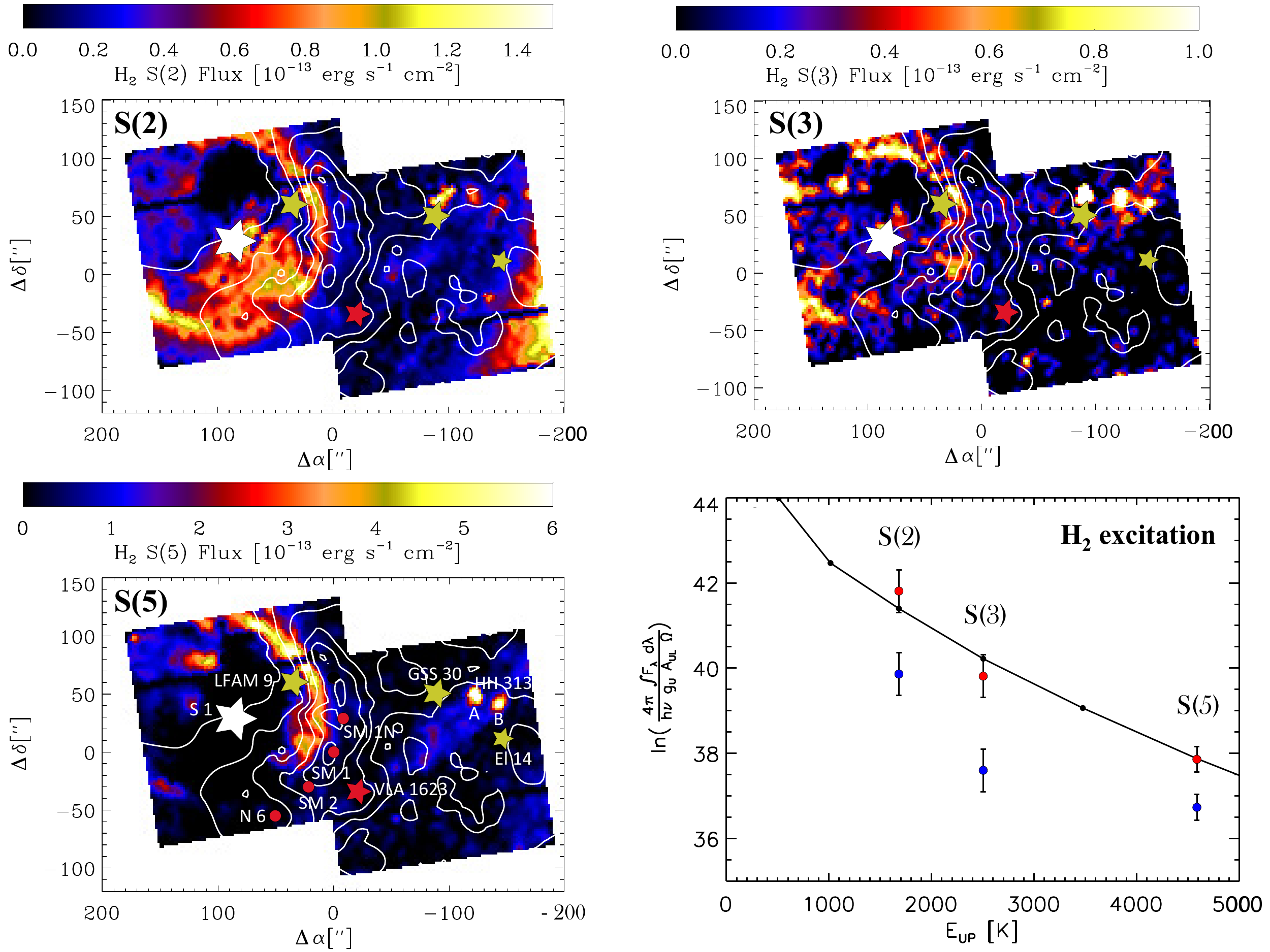}
 \caption{{\bf Upper panels:} (from left to right) Distribution of the emission in \oishort, \oilong,\ and the ratio of their intensities $I$(\oishort)/$I$(\oilong). The illuminating source, S\,1, is depicted as a white star on the left side of the frames, being at the centre of the dashed circles. The white contours outline integrated \ntwohp\,(3-2) emission.
  {\bf Lower panels:}   Distribution of the S(2), S(3), and S(5) pure rotational lines of \molh\ are shown by the coloured maps, with the fluxes given by the scale bars. These data were obtained with the ISOCAM-CVF. The white contours and the symbols are as in Fig.\,\ref{obs_outlay}. The graph shows the fit with the Meudon PDR code \citep[][red dots]{lepetit2006} to the dereddened observations for \av\,= 6\,mag. Blue data points are the observed values. Other \molh\ lines, i.e. S(1) and S(4), are also filled in and shown as black dots.   
        }
  \label{h2}
\end{figure*}

%
%

\section{Observations and data reduction}

\subsection{\rm \water, \ammonia,\ and \ntwohp}

%
%

The \roa\ cloud was observed in two \water\ lines with the Heterodyne Instrument for the Far Infrared \citep[HIFI,][]{degraauw2010} on board the ESA spacecraft {\it Herschel} \citep{pilbratt2010}  carrying a 3.5\,m telescope. The 557\,GHz data were acquired during OD\,475, 1226, and 1231 in mixer band 1b and the 1670\,GHz (179.5\,\um) data on OD\,1221 in band 6b. Data were taken simultaneously in the H- and V-polarizations using both the Wide Band Spectrometer (WBS, accousto-optical) with 1.1\,MHz resolution and the High Resolution Spectrometer (HRS, correlator) with 0.25\,MHz resolution. The Band\,1 receiver was tuned such that rotational transitions of ammonia and diazenylium, i.e. both the \ettnoll\,572\,GHz line and the \ntwohp\,(6-5)\,560\,GHz transition, were also simultaneously admitted. The map and observational details for diazenylium can be found in Paper\,I.

For the programme  GT2\_rliseau\_1, the mapping was done on the fly (OTF) with a reference position at  ($-10$\amin, $-10$\amin) on OD\,475, 1226, and 1231. At 557\,GHz, the map size was 8\amin\,$\times$\,\amindot{6}{5} with samplings in 16\asec\ steps. At 1670\,GHz (OD\,1221), the map was \amindot{2}{5}\,$\times$\,1\amin\ with a sampling rate of 20\asec. The observing times were 7 and 3 hours, respectively.  All data have been reduced with the HIFI pipeline HIPE version 14.1.0. Instrument characteristics are provided by \cite{roelfsema2012} and some of the observational parameters can be found in Table\,\ref{Herschel}.

The \ettnoll\ and \ntwohp\,(6-5) lines were simultaneously  observed with the \water\,($1_{10}-1_{01}$) transition. These rectangular maps were all observed at an angle of 55\adeg\ with respect to the equatorial coordinate system and had to be rotated for proper alignment. The rotation and regridding data can be found in \citet{bjerkeli2012} and in Paper\,I.
 
\subsection{\rm O$_2$}

Following up on our initial \molo\ observations with HIFI \citep{liseau2012}, we  obtained data in the  \molo\,($3_3-1_2$)\,487\,GHz line towards 12 new positions on OD\,1357, 1358, 1367, and 1371 (OT2\_rliseau\_2, 38.5\,hr integration time).  In addition, the \molo\,($5_4-3_4$)\,773\,GHz line in HIFI Band\,2 was observed towards HH\,313\,A, an optically visible shocked region of the VLA 1623 outflow (8.5\,hr integration on OD\,1384). The reduction of these new data also followed the procedures outlined in the 2012 paper. 

All in all, in \roa\ we observed \molo\,487\,GHz towards 15 positions and \molo\,773\,GHz towards five highly oversampled positions, each corresponding in size to that of the 487\,GHz data. The 773\,GHz {\it Herschel} beam  was 27\asec\ and the grid spacing 10\asec.

\subsection{\rm O\,I}

Observations of the fine structure components of O\,I at 63\,\um\ (\pone-\ptwo) and 145\,\um\, (\pzero-\pone) of the inverted ground state with the ISO-LWS \citep[Long Wavelength Spectrometer,][]{clegg1996} towards the VLA\,1623 outflow have been presented earlier by \citet{liseau2009}. On OD\,1202, the Photodetector Array Camera and Spectrometer \citep[PACS,][]{poglitsch2010} was used to map the fine structure line emission north of the outflow (see Fig.\,\ref{obs_outlay}) as part of the programme GT2\_rliseau\_2 and with an observing time of 2\,hr 16\,min. The map consists of $3 \times 4$ PACS rasters separated by 38\asec. An individual PACS raster covers 47\asec\,$\times$\,47\asec\ and is composed of $5\times 5$ spaxels (spatial pixels), providing a spatial sampling of \asecdot{9}{4} per spaxel. At 63\,\um, the spectral resolution is about 3500, and at 145\,\um\ \about\,1000, corresponding roughly to 100 and 300\,\kms, respectively. The lines in \roa\ are therefore expected to be unresolved, except perhaps towards the outflow. 

The observing mode was {\it unchopped grating scan} and a position about \adegdot{1}{5}\,NE of the map was used as a reference. To estimate the absolute accuracy of the fluxes we compared data from one map position with those that we previously obtained with the ISO-LWS  \citep[e.g.][]{liseau2009}. However, we refer here to re-reduced data using the final version of the pipeline (OLP\,10). The LWS line and continuum fluxes were both measured using two methods:   the default point-source calibration and  by exploiting extended source corrections. To compare the PACS measurements with those with the LWS, the PACS data were  convolved with a flattened top Gaussian beam \citep{larsson2000} and with a pure equivalent aperture measurement. The results show that these two different methods lead to intra-instrumental agreement within 10-30\%.

\subsection{\rm \molh\ and polycyclic acromatic hydrocarbons}

Towards  \roa, the Infrared Space Observatory (ISO) archive contains three frames of spectrophotometric data using the ISOCAM-CVF \citep[Circular Variable Filter,][]{cesarsky1996}. For one frame, the pixel field of view (FOV) is 300\asec\ and it images the region around the star GSS\,30. The other two, with a pixel FOV of 600\asec, cover the VLA\,1623 outflow \citep{liseau2009} and the PDR of S\,1, respectively. Together, these two frames cover most of the \roa\  region. In addition, {\it Spitzer} data of the NW part of the VLA 1623 outflow have been presented by \citet{neufeld2009}.

Towards the S\,1 PDR region, PAH emission totally dominates the spectrum over the wavelength range of the ISO-CVF \citep[see also][]{justtanont2008}. However, three pure rotational \molh\ lines,   S(2), S(3), and S(5),  could also be extracted. The S(5) line at 6.9\,\um\ is relatively unaffected by other spectral features, while the S(2)\,12.3\,\um\ line is blended with a PAH feature and the S(3)\,9.3\um\ line is situated inside the deep silicate absorption. Consequently, the fluxes for these two lines are more uncertain. The observations and procedures to reduce the CVF data are described in detail by \citet{liseau2009}. There, numerical model fits to the spectral PAH features were used to estimate the fluxes also of other \molh\ lines. Comparing these results with the present ones, where possible, demonstrates excellent agreement.

\subsection{CS and CO}

CS\,(10-9) data were obtained simultaneously  with the \molo\,487\,GHz maps with HIFI. Complementary data had previously been collected with the 15\,m Swedish ESO Submillimetre Telescope \citep[SEST,][]{booth1989} in the CS transitions of (2-1), (3-2), and (5-4), in addition to pointed observations in the isotopologues C$^{34}$S and $^{13}$C$^{34}$S (see Appendix\,C). 

Simultaneously with the \molo\,773\,GHz line, $50^{\prime \prime} \times 50^{\prime \prime}$ maps at 10\asec\ sampling of  $^{13}$CO\,(7-6) and C$^{17}$O\,(7-6) were obtained (see Fig.\,\ref{CO_iso} below).

\begin{figure*}
 \resizebox{\hsize}{!}{
  \rotatebox{00}{\includegraphics{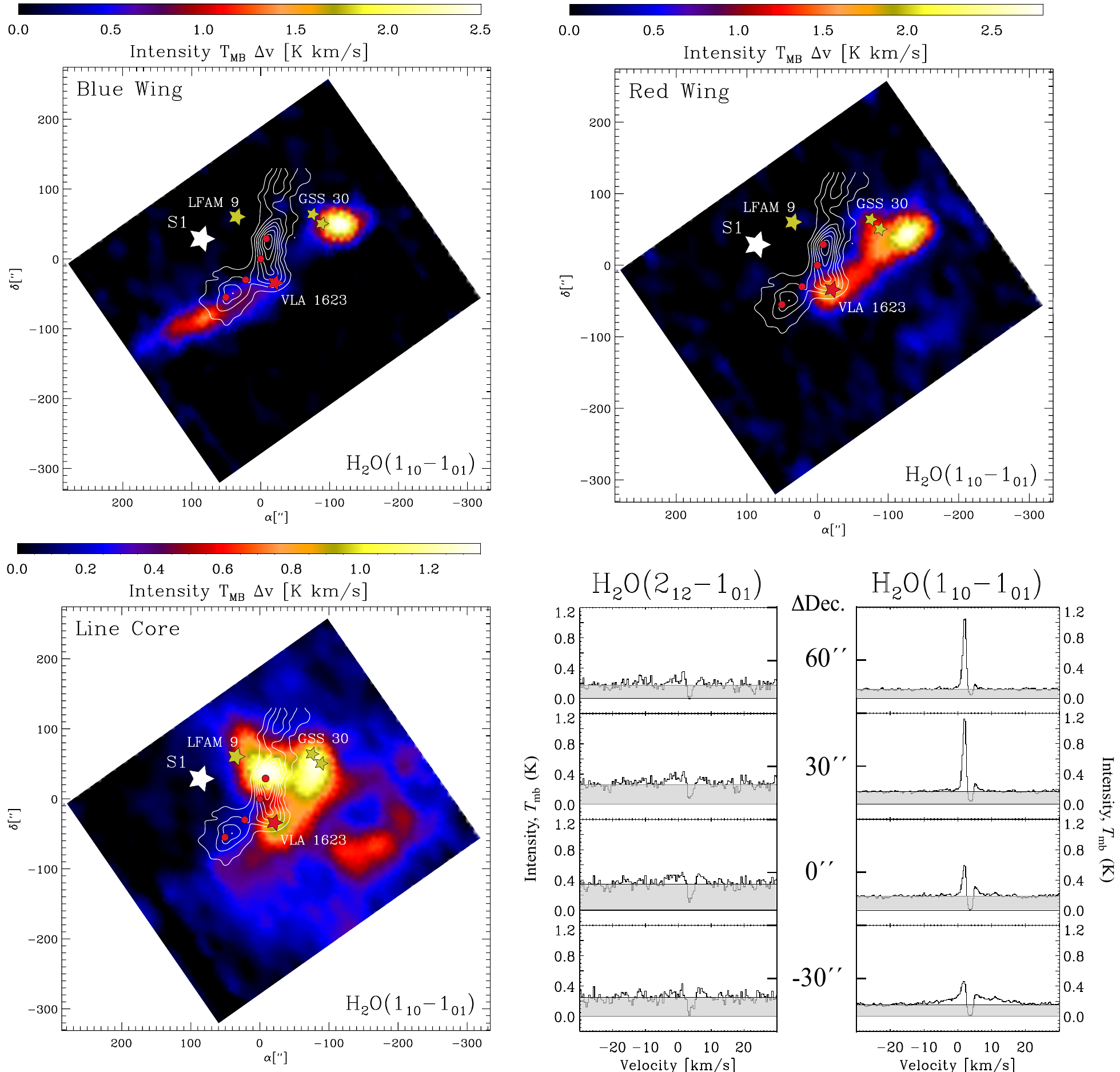}}
  }
  \caption{{\bf Upper and lower left:} HIFI maps of the kinematic components of the ortho-\water\,557\,GHz line referring to the VLA\,1623 outflow, labelled `Blue Wing' and `Red Wing', respectively, and that of the quiescent gas of the `line core'. Specifically, the integrations of the intensity, \tdv, were  over [$-30,\,-2$[\,\kms\ for the blue wing and  over ]$+5,\,+40$]\,\kms\ for the red wing.The line core is bounded within [$-2,\,+2$]\,\kms.  A deep absorption feature is observed in the range \vlsr = [$+2,\,+5$[\,\kms. The contours  outline observed \ntwohp\,(3-2) emission.
{ \bf  Lower right:}  HIFI north-south strip map about the Right Ascension offset $\Delta \alpha =  0$\asec\ in \water\,1670\,GHz (left) and 557\,GHz (right), see also Fig.\,\ref{obs_outlay}. The declination offsets $(\Delta {\rm Dec.)}$ are indicated  between the frames. Both data sets were convolved to a beam size of 38\asec. To compensate for the observation in double sideband mode the continuum intensity, shown in shaded grey, has been divided by a factor of two. 
        } 
  \label{h2o_comps}
\end{figure*}

\begin{figure*}
  \resizebox{\hsize}{!}{
    \rotatebox{0}{\includegraphics{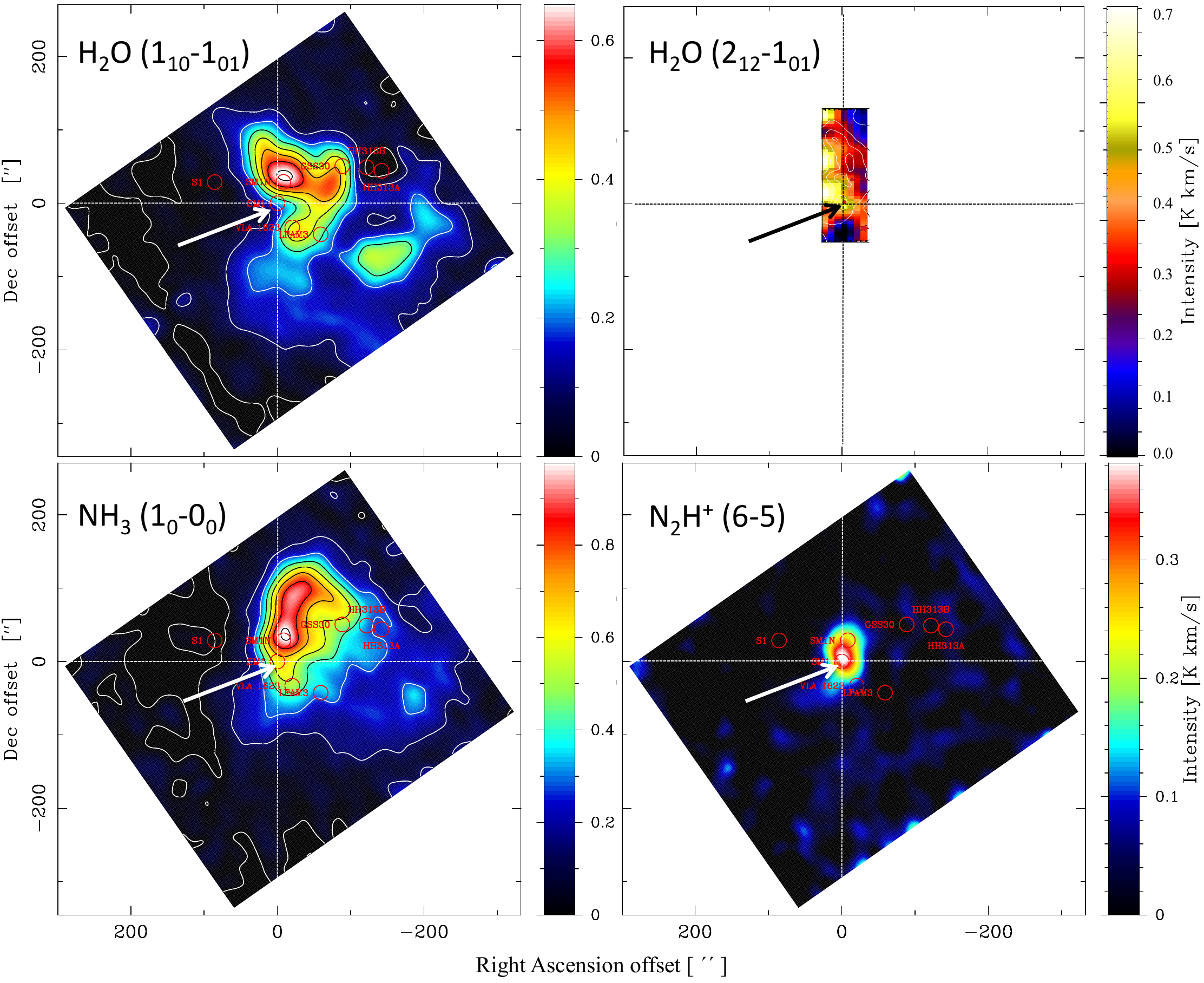}}      
                        }
  \caption{{\it Herschel}-HIFI maps of integrated water, ammonia, and diazenylium emission. The scale bar to the right of each frame is in K\,\kms\ and the arrows all point to the location of the dark core SM\,1 at the crosshair. Neither \ammonia\ nor \water\ peaks there, but \ntwohp\  does. The maps of the line spectra of \ntwohp\,(3-2) and (6-5) are presented in Paper\,I.
  }
    \label{nh3_h2o_n2h+}
\end{figure*}

\begin{figure*}
  \resizebox{\hsize}{!}{
    \rotatebox{0}{\includegraphics{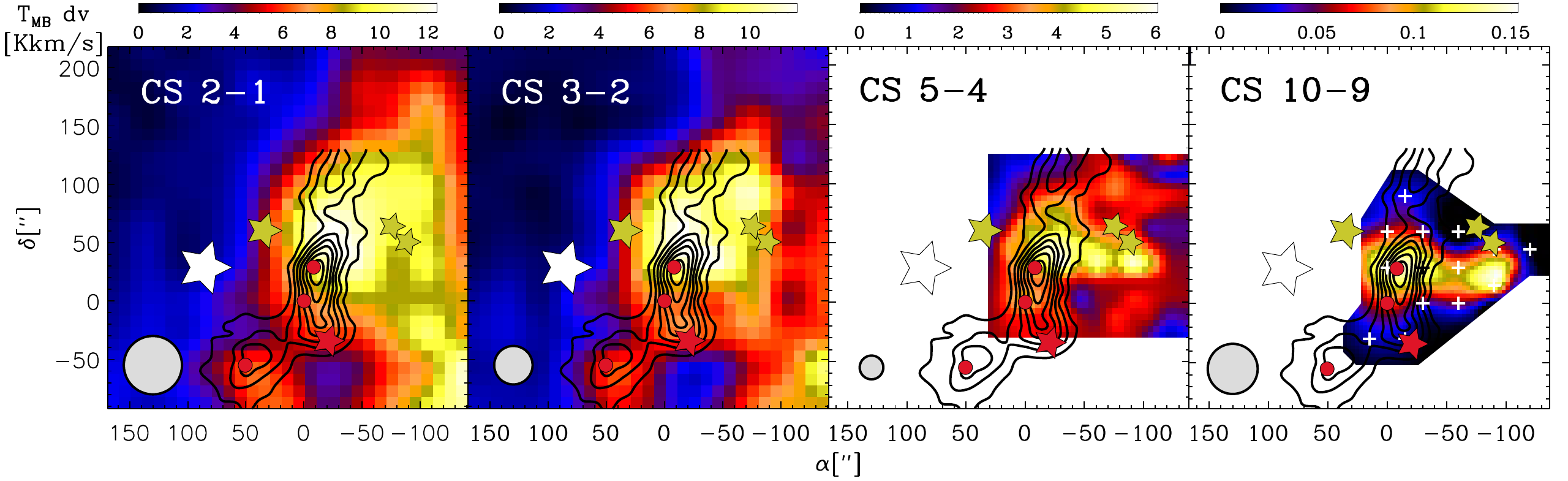}}  
                        }
  \caption{Maps of integrated CS line intensity, \tmbdv, of CS\,(2-1), (3-2), (5-4), and (10-9). Symbols are as in Fig.\,\ref{obs_outlay} and the half-power widths of the telescope beams are shown as grey circles in the lower left corners (see Tables\,\ref{Herschel} and \ref{SEST}). The contours show the distribution of \ntwohp, as traced by its (J=3-2) line (Paper\,I).
  }
    \label{CS_1}
\end{figure*}

\begin{figure*}[ht]
  \resizebox{\hsize}{!}{
  \rotatebox{00}{\includegraphics{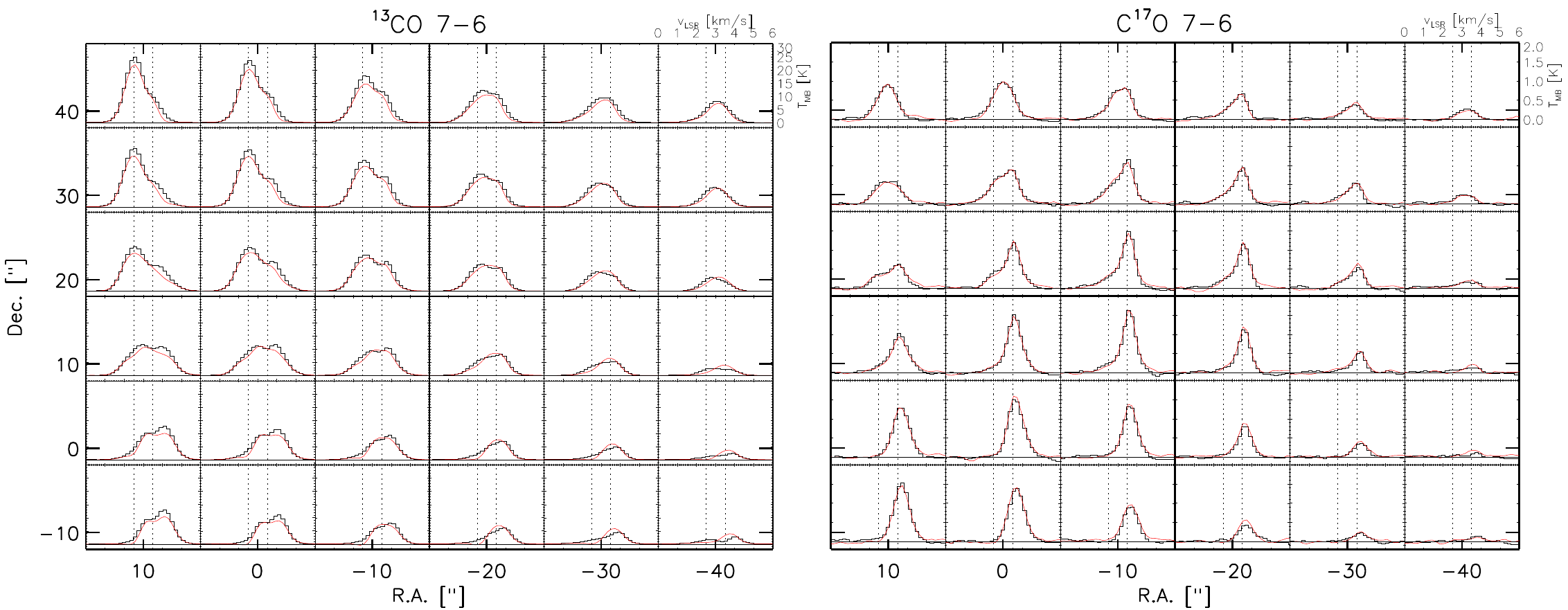}}
  }
  \caption{Oversampled maps in $^{13}$CO\,(7-6) (left) and C$^{17}$O\,(7-6) (right) with the 27\asec\ beam of {\it Herschel}. The \vlsr\ and \tmb\ scales are indicated in the upper right corners. The dashed vertical lines, at LSR velocities of 2.5\,\kms\ and 3.5\,\kms, respectively, identify two radial velocity components of the fitted Gaussians, shown by the smooth red curves, whereas the observations are shown as histograms.
        }
  \label{CO_iso}
\end{figure*}

\begin{figure*}
  \resizebox{\hsize}{!}{
  \rotatebox{00}{\includegraphics{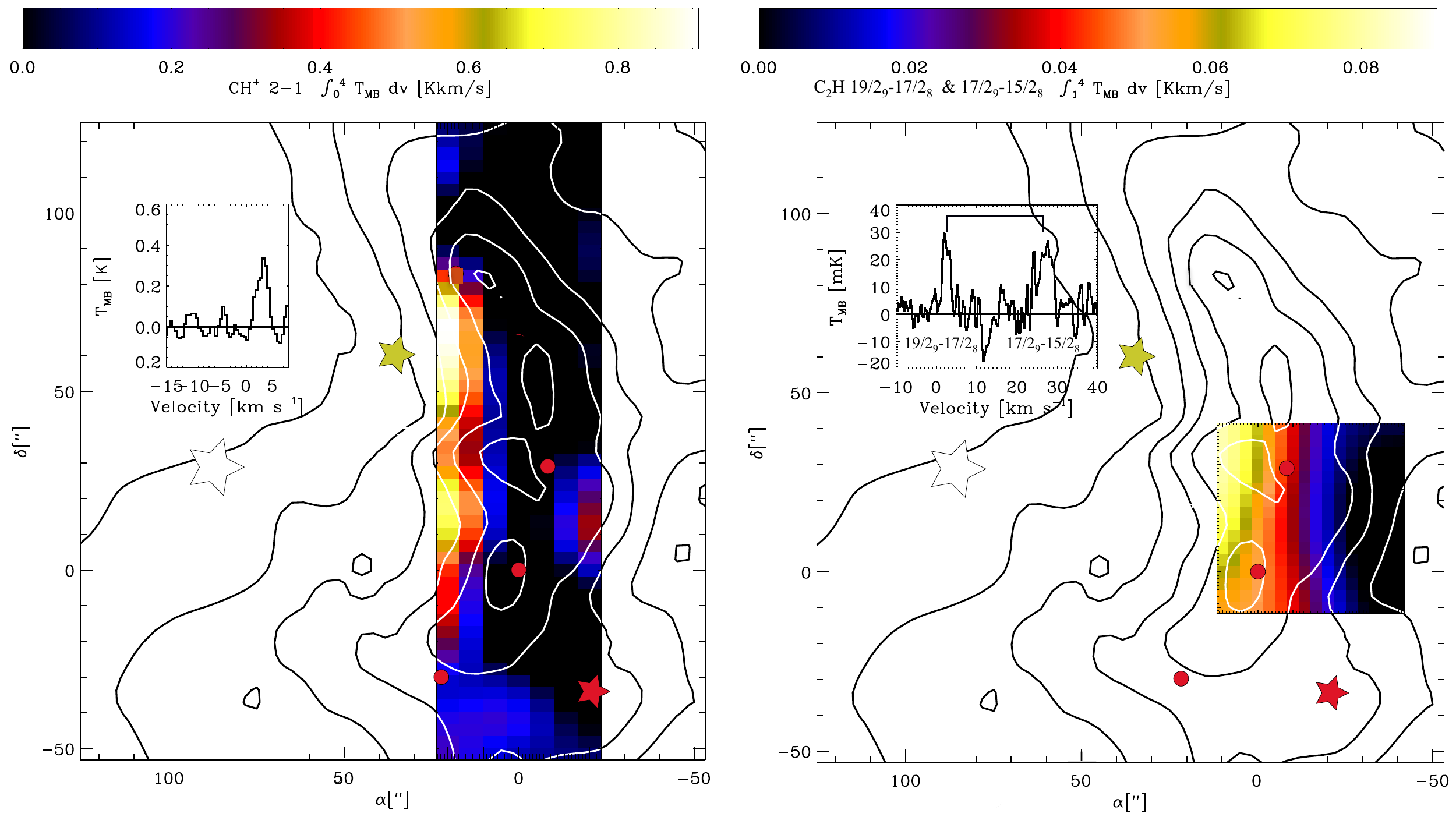}}
  }
  \caption{{\bf Left:} HIFI-map of the 1.67\,THz methylidine line, CH$^+$\,(2-1), obtained simultaneously with \water\,($2_{12}-1_{01}$).
  {\bf Right:}  Small map of the emission in ethynyl, C$_2$H\,$(17/2_9-15/2_8, 19/2_9-17/2_8)$\,786\,GHz, observed together with \molo\,($5_4-3_4$)\,773\,GHz. 
  The contours and symbols are as in Fig.\,\ref{obs_outlay}.
        }
  \label{C2H}
\end{figure*}



\section{Results: A spatially resolved PDR}

The mapping in atomic and molecular line emission revealed a spatially resolved PDR, seen from the side (`edge-on'). 

\subsection{Hydrogen, \molh}

As can be seen in Figure\,\ref{h2}, the \molh\ S(2) emission outlines a clear spherical shell structure of warm PDR gas around the stellar object S\,1. In the figure this structure is also seen in the S(3) and S(5) lines, but at a slightly more fragmented level. This spherical structure motivates the radial averaging of the \molh\ fluxes and these are plotted in the lower rightmost frame of Fig.\,\ref{h2}, together with a model fit using the Meudon PDR code\footnote{Unless explicitly noted, we are using standard values of the code.} \citep[][see below]{lepetit2006}. The formation of \molh\ on grain surfaces is taken into account \citep{lepetit2009,lebourlot2012}.

The observed data are consistent with gas at a temperature of about \powten{3}\,K that has a total column density of  a few times \powten{19}\,\cmtwo, values that are identical to those determined by \citet{liseau2009} towards the VLA\,1623 outflow, and where an \molh\ ortho-to-para ratio of 2 was derived. 

\subsection{Oxygen: O\,I and \molo}

\subsubsection{Atomic oxygen, O\,I}

As in the \molh\  image, the VLA\,1623 outflow is also seen in the upper part of the leftmost panel of Fig.\,\ref{h2}, showing the distribution of the fine structure line emission of [O\,I]\,(\pone-\ptwo)\,63\,\um. In the upper middle panel,  the distribution of the (\pzero-\pone)\,145\,\um\ line, originating from a higher upper level\footnote{The upper level energy $E_0/k$ of the 145\,\um\ line is 326\,K above ground, and for the 63\,\um\ transition, $E_1/k=228$\,K.}, is also shown. These atomic fine structure lines mark very clearly the boundary seen in \molh, but also show significant emission throughout the entire molecular cavity around the stellar object S\,1. As seen from the centre to the edge of this region, the extinction is limited to $A_{\rm V} \le 3 \times 10^{-3}$ magnitudes, equivalent to a hydrogen column density of $4.2 \times 10^{18}$\,\cmtwo\ (see Paper\,I).  For a solar oxygen abundance \citep[$5 \times 10^{-4}$,][]{asplund2009} we would expect an [O\,I]\,63\,\um\ flux of about $4 \times 10^{-12}$\,\ecs. Observed flux levels are lower than $6\times 10^{-13}$\,\ecs. In a low-density medium and at temperatures of some \powten{3}\,K, essentially all O$^0$ atoms are in the \ptwo\ ground state. This observed flux deficiency could therefore indicate that the 63\,\um\ line is self-absorbed. This hypothesis can be tested by a comparison with the observed 145\,\um\ transition. 

The upper right panel in Fig.\,\ref{h2} displays the line flux ratio $F_{63\,\mu {\rm m}}/F_{145\,\mu {\rm m}}$. In PDRs, observed values are generally around 10 to 20 \citep{hollenbach1999}. It is therefore remarkable that some regions show ratios ten times lower, and even lower than unity. The region in the southeast where this occurs, i.e. N\,6, is otherwise conspicuous only in \ntwohp\ emission \citep[][and Paper\,I]{difrancesco2004}. 

Line ratios of the magnitude observed here are characteristic of gas that is neither optically thin nor in thermodynamic equilibrium \citep[e.g.][]{liseau2006,canning2016}. Oxygen fine structure lines from cold gas (\lapprox\,20\,K), which are  optically thick in both ground state transitions, can show ratios below unity. A gas kinetic temperature of 17\,K in this particular region was recently determined  in Paper\,I. However, the N\,6 column density, $N({\rm H}_2)=2\times 10^{23}$\,\cmtwo, is too low by a factor of 300 for the \pzero$_{-1}$\,145\,\um\ optical depth to reach unity \citep[see][]{liseau2006} and must therefore be dismissed as an explanation. Instead, as advocated by these authors, a cold foreground cloud that absorbs much of the \pone$_{-2}$\,63\,\um\ radiation but leaves the higher level \pzero$_{-1}$\,145\,\um\ line unaffected appears to be the only viable alternative to explain the very small 63\,\um\ to 145\,\um\ flux ratios of N\,6.

\subsubsection{Molecular oxygen, \molo}

In Fig.\,\ref{h2} it can be seen that the atomic oxygen, O\,I, appears to be penetrating quite deeply into the dense parts of the dark core \roa, outlined by the contours of the \ntwohp\ emission. It feeds the \molo\ production region (Fig.\,\ref{o2_map}) with atomic oxygen. There, the molecular oxygen is produced in the gas phase, with additional contribution from the ice-crusted dust grains \citep{hollenbach2009}. The models of \citet{melnick2015} for Orion\,A, which invoke shock chemistry, do not likely  apply to \roa\ as the \molo\ emission regions and the outflow/HH-shocks are spatially well separated. In \ro, these shocks enhance the \water\ abundance, but not that of \molo. 

Unfortunately, our \molo\ map is not complete and maximum emission occurs near the edge of the map. To remedy this, complementary observations would be required. However, after the decommissioning of {\it Herschel}, there is no other facility available or planned any time soon. The factor of two higher intensity at the local \molo\  emission maximum is likely an effect of the increased temperature (25\,K) rather than a higher column density of \molo\ molecules.

\subsection{Water, \water}

\subsubsection{Spectral components of the line profiles}

The emission due to water is seen essentially everywhere in the molecular core \roa, except in the easternmost parts towards the early-type stellar object S\,1, where the water molecules have been destroyed by the radiation from the B-star. 

The lines of the two ground state transitions\footnote{For the lowest rotational levels, an energy diagram can be found in \citet{ewine2011}.} of ortho-\water\ have a different appearance. For  the ($1_{10}-1_{01}$)\,557\,GHz line, the spectral profiles exhibit a complex structure with emission peaks, deep central absorptions, and extended wings. Maps of these features reveal that these dominate different parts of the cloud core (Fig.\,\ref{h2o_comps}). That is, the wings outline the bipolar, jet-like molecular outflow from the protostellar object VLA\,1623 \citep{bjerkeli2012}, whereas the line core is most prominent in the denser parts of \roa\  where $n$(\molh)\,$\ge 3 \times 10^6$\,\cmthree\ (paper\,I). Comparing the 557\,GHz HIFI data with those that were obtained earlier with Odin (see Appendix\,A) and also SWAS\footnote{Submillimeter Wave Astronomy Satellite \citep[SWAS,][]{melnick2000}.} shows that these components of the line shape are excellently reproduced by different instruments in independent observations (Fig.\,A.2).

The HIFI observation of the higher excitation line \water\,($2_{12}-1_{01}$)\,1670\,GHz is a completely new result, but with its three times smaller beam the map is not as extended as that for the 557\,GHz line. However, our strip map of the dense core region of \roa\ (Fig.\,\ref{obs_outlay}) is still useful for a comparison of the two line profiles (see Figure\,\ref{h2o_comps}).  The 557\,GHz line displays several components, whereas the 1670\,GHz line is mostly in absorption with weak wing emission towards the dark cores SM\,1 and SM\,1N. This may seem surprising, as these positions are not anywhere near the strong outflow from VLA\,1623 (Fig.\,\ref{h2o_comps}), but could be due to the weak CO outflows of \citet{kamazaki2003}.

The most significant part of the line profile, though, is the deep absorption feature, reaching down to close the zero level. The same depth is also seen in the $(1_{10}-1_{01})$ line, where the signal-to-noise (S/N) is significantly higher. There, however, the emission core varies appreciably in intensity with position. This is not an effect caused by a foreground absorber drifting over the cloud, thereby changing the strength of the emission peak, because the centre of the absorption dip in both lines is completely stable with respect to the systemic Velocity Standard of Rest (\vlsr).

\subsection{Ammonia, \ammonia}

The spatial distribution of the \ettnoll\ line is similar to that of \water\,($1_{10}-1_{01}$), both being of the ortho flavour. As for water and diazenylium (\ntwohp), ammonia peaks close to SM\,1N. A clear difference is displayed by deuterated species, which all peak at the SM\,1 position. For instance,  formaldehyde has its emission maximum towards SM\,1N, whereas its doubly deuterated form peaks towards SM\,1 \citep{bergman2011a}, i.e. D$_2$CO has essentially vanished towards SM\,1N where the H$_2$CO emission is strongest. This difference may have its explanation in the fact that temperatures are lower at SM\,1 (Paper\,I), promoting the exothermic process of molecular deuteration. However, \citet{friesen2014} detected H$_2$D$^+$ towards SM\,1N, but not towards SM\,1.

The ground state lines of both \water\ and \ammonia\ have double-peaked line profiles. However, a remarkable difference between the ammonia and water lines are the high-velocity wings that are entirely absent in the ammonia data (Fig.\,\ref{nh3_profiles}). Consequently, low-amplitude velocity fields like those due to gravitational infall are potentially better traced in the \ammonia\ lines.


\subsection{Carbon monoxide, CO}

In an adjacent shell, next to \molh, low-$J$ CO is distributed along a dense ridge that is broken up into several high-density clumps, of which SM\,1 and SM\,1N  are the most conspicuous ones (Fig.\,\ref{obs_outlay}).
 
Profile maps of  $^{13}$CO\,(7-6) and C$^{17}$O\,(7-6) have been secured simultaneously with the observations of \molo\,($5_4-3_4$)\,773\,GHz as prime. The positions are identified in Fig.\,\ref{o2_map} by the red rectangles. The profiles are fit by a two-component model, where Gaussians were fitted to the observations. Westward of the ridge, the isotopologues of CO decline rapidly.

\subsection{Methylidine, CH$^+$, and ethynyl, C$_2$H}

At the very edge of the PDR-interface, the ($J=2-1$) line of the cation CH$^+$ has been mapped simultaneously with the \water\,($2_{12}-1_{01}$) line at 179.5\,\um. In the left frame of Fig.\,\ref{C2H}, it can be clearly seen that the emission arises in front of the cores SM\,1N, SM\,1, and SM\,2 (red dots from north to south) as seen from the B star S\,1. 

In contrast, the neutral C$_2$H radical is detected further in and closer to the cores, where the UV penetration into the cloud has diminished due to the increased extinction by the dust (Fig.\,\ref{C2H}, right-hand panel).

\subsection{Carbon sulfide, CS}

The CS maps, up to ($J=5-4$), were obtained at the SEST (Appendix\,C) and the ($J=10-9$) data with HIFI aboard {\it Herschel}. These maps indicate, following the excitation gradient of the PDR, that CS appears behind CO and not predominantly in the densest parts of the cloud core.


\section{Discussion}


\subsection{The \roa- PDR}

For the stellar parameters given in Table\,5 of Paper\,I, and assuming solar elemental abundances, we use ATLAS\,9 model atmospheres for the energy distribution of the stars \citep{castelli2004}.  At 0.05\,pc from S\,1, the FUV field\footnote{For a definition of $G_0$, see e.g. \citet{hollenbach1999}.} amounts to roughly $5000$ in terms of $G_0$, whereas the western side is illuminated by a $G_0 \sim 100$ field from HD\,147889. 

The extinction towards \roa\ is well described by a law with \rv\,=\,5.5, so that the gas-dust relation reads $N$(H\,I + \molh)$\,=N$(H\,I)$+ 2\,N$(\molh)$ =1.4 \times 10^{21}$\,\av\,\cmtwo\  \citep[][see also Paper\,I]{bohlin1978}. 

\subsection{The physical conditions in \roa}

\subsubsection{A gaseous sphere around S\,1}

At distances greater than $3 \times 10^{16}$\,cm from the centrally positioned S\,1, radiation pressure on the hydrogen gas is not very important. Due to the efficient braking by the 600 to \powten{3}\,\cmthree\ gas (seen in [O\,I] throughout this region, Fig.\,\ref{h2}), radiation pressure in the Lyman lines can result in terminal velocities of only a few times \powten{-2}\,\kms\ at the 0.05\,pc interface\footnote{These computations  are based on an ATLAS\,9 model atmosphere for $T_{\rm eff}= 17\,000$\,K, $\log\,g = 4.0$, [M/H] = 0.0 \citep{castelli2004}.}. This is much smaller than the turbulent speed of a few times \powten{-1}\,\kms\ in the cloud core and  therefore has   little impact on its dynamics.

Without any significant dynamical pressure gradients, the nearly perfectly spherical shape of this region is kept in place by a delicate pressure balance, where the higher temperature of the H region is offset by the higher density of the molecular core, i.e. $P_{\rm H} / P_{\rm H_2}  \sim  600\,{\rm K} \times 3\,10^4\,{\rm cm^{-3}} / 10\,{\rm K} \times 2\,10^6\,{\rm cm^{-3}}$.  In high-($G_0/n$) PDRs the situation is different, where an H\,II region drives an ionization front into the molecular cloud \citep[e.g.][]{tielens2005}. A clumpy structure would further alter both molecule formation depths and, in particular, the radiative transfer.  

\subsubsection{Abundant \molo\ production in \roa}

Of considerable interest is to understand the formation history of \molo\ in \roa\  as this molecule, in spite of numerous attempts \citep{goldsmith2000,goldsmith2002,pagani2003,yildiz2013,sandqvist2015,wirstrom2016}, has 
hardly been found anywhere else outside the solar system. Taking into account new lab results for the oxygen binding energy on dust grains \citep{he2015}, \citet{taquet2016} recently discussed a number of formation scenarios and argued in favour of a particular one that could fit the physical conditions in \roa.

\citet{taquet2016} suggested that the high abundance of \molo\ seen in the gaseous comae of two solar system comets, with $X$(\molo)/$X$(\water) = 0.01-0.1 \citep{bieler2015,rubin2015}, points towards a primordial origin, i.e. that this \molo\ was initially produced in the protosolar cloud core and subsequently transported through the viscous protoplanetary disc to its inner regions. It is argued that chemistry on icy dust grains would produce observed levels of related species, i.e., $X$(H$_2$O$_2$)/$X$(\molo) and $X$(HO$_2$)/$X$(\molo), see \citet{bergman2011a} and \citet{parise2012}, respectively, and with the accompanying theoretical chemistry models presented by \citet{du2012a, du2012b}.

However, we find it very difficult to reconcile the comet results for \molo\ and \water\ with our abundance determinations for \roa, i.e. $X$(\molo) = $5 \times 10^{-8}$ \citep[][this paper]{larsson2007,liseau2012} and $X$(\water) = $5 \times 10^{-9}$ (see  Sect.\,4.3.2), hence $X$(\molo)/$X$(\water) = 10. 

In a model specifically designed for \roa, \citet[][see their Fig.\,6]{taquet2016} find that $X$(\molo)/$X$(\water)$\,> 1$ during the time of about 5000 -  30\,000\,yr on their chemical clock, with a free-fall time of 16\,000\,yr (see below) falling right into this interval. If this chemical model is basically correct (there might be issues with their HO$_2$ and H$_2$O$_2$ abundances), the data would suggest the dense clumps in \roa\ to be (chemically) very young indeed. As discussed by \citet{liseau2012}, the relative brevity of the period of abundant \molo\ in the gas phase would readily explain the elusiveness of this molecule in the interstellar medium.


\subsection{Theoretical models of infall: \water}
 
 The HIFI instrument aboard {\it Herschel} is based on the heterodyne technique that implies high spectral resolution, $\nu/\Delta \nu > 10^6$. At the operating frequencies of HIFI, radial velocity resolutions are fractions of a \kms,  sufficient to spectrally resolve narrow molecular lines that originate in the cold interstellar medium.
 
 \subsubsection{Line profiles: \water\,($1_{11}-1_{10}$)}
 
Due to their expected high optical depths ($>100$), the ground state lines of water are particularly suited as tracers of protostellar infall, as these large optical depths lead to large contrasts between receding and approaching emission. The high optical depths, in combination with the complex system of its many energy levels of the water molecule, imply however that the radiative transfer is difficult and the proper computation is therefore often avoided. A widely used method is to fit the observed profiles, essentially `perfectly', with multi-component Gaussian functions; however, this procedure does not  appropriately recover the physics and so we turn  to other options.

We compute the line profiles from a physical model and account for the radiative transfer using an Accelerated Lambda Iteration (ALI) code; at these high optical depths other possibilities such as Monte Carlo methods frequently 
run into conversion problems. The benchmarking of the ALI code has been described by \citet{maercker2008}. In the present work, the collisional rate constants of \cite{faure2007} are used.

\subsubsection{Profiles of the infall centre}

The observed 557\,GHz line profile, with its blue-red asymmetry and deep central absorption, is highly reminiscent of theoretical profiles of spherical infall, i.e.  from a protostar at the onset of gravitational collapse prior to observable disc formation \citep[e.g.][]{ashby2000}.

Because of its conceptual and computational simplicity \citep[see e.g.][]{foster1993}, our protostar model is a Bonnor-Ebert sphere \citep[BE,][]{bonnor1957,ebert1957};  based on observed criteria (Paper\,I), it is found to be gravitationally unstable (dimensionless mass\footnote{$m=(4\,\pi\,\rho_{\rm 0}/\rho_{\rm out})^{-1/2}\,(\xi^2\,d\psi/d\xi)_{\xi_0}$, where $\psi$ and $\xi$ are the dimensionless gravitational potential and the radial variable in the Lane-Emden equation, respectively.} = 14.1, density contrast $\rho_{\rm 0}/\rho_{\rm out}= 29$).  The critical BE mass is 1.04\,\msun\ within 3000\,AU (25\asec), where the thermal pressure is $1.24 \times 10^{-9}$\,erg\,\cmthree\ ($T_{\rm out}=30$\,K, $n_{\rm out}=3 \times 10^5$\,\cmthree)  and a free-fall time is 16\,000\,years. 

The line profiles to be compared to the observed ones are computed self-consistently, based on observation, and with the following ALI  parameters: $T_{\rm kin}=9$ to 15\,K, $\upsilon_{\rm turb}=0.4$ to 0.6\,\kms,  gas-to-dust mass ratio of 17.5, and a dust-$\beta=1.8$.  We found an average water abundance $X$(\water)\,$= 5 \times 10^{-9}$ that resulted in a central optical depth of 160 in the 557\,GHz ground state line. This model correctly reproduces the observations of the spectra at the central position in both the ground-state lines of ortho-\water, i.e. the $(1_{10}-1_{01})$ 557\,GHz and the $(2_{12}-1_{01})$\,1670\,GHz transitions.

\begin{figure*}
 \resizebox{\hsize}{!}{
  \rotatebox{00}{\includegraphics{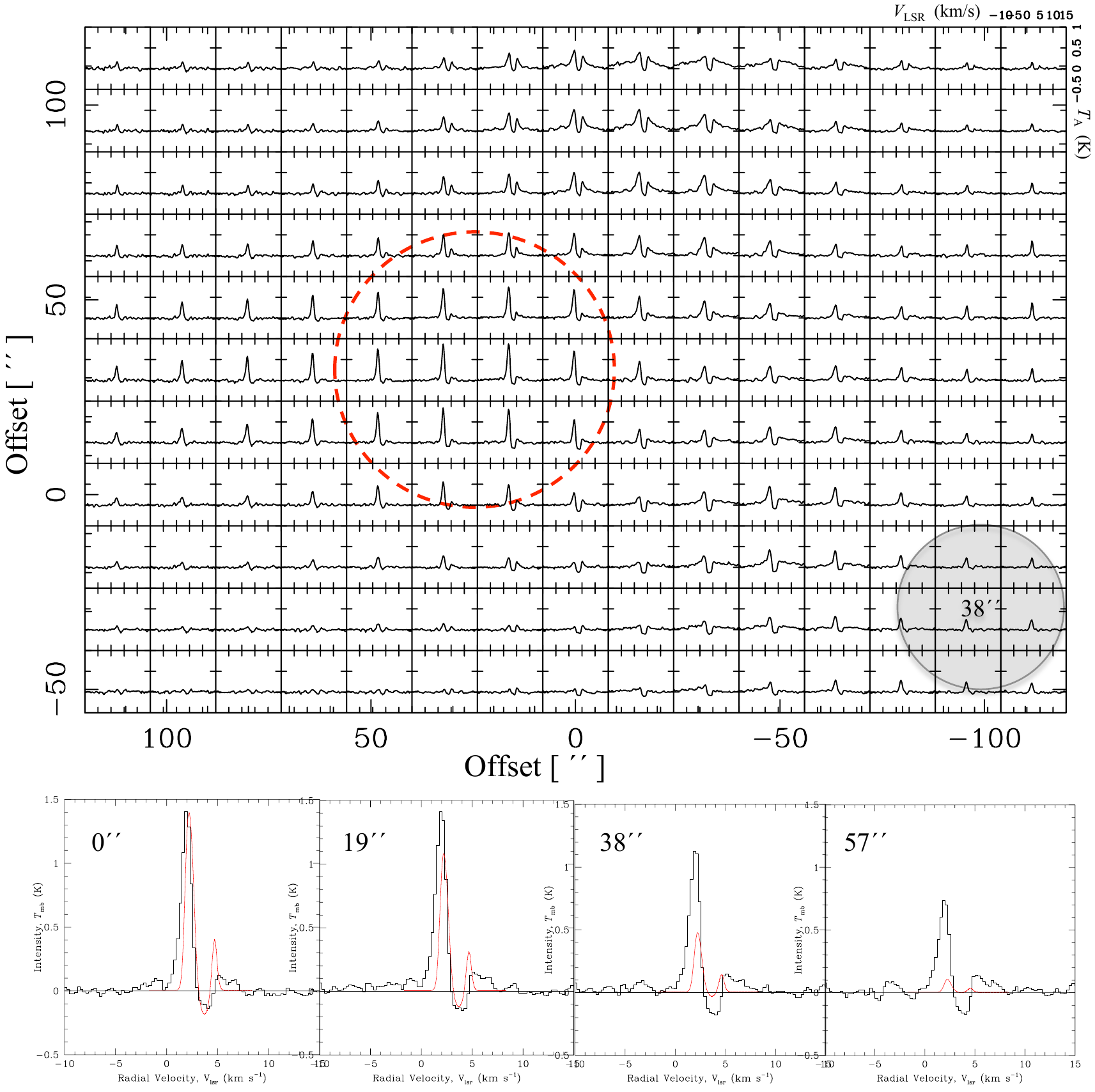}}
  }
  \caption{{\bf Upper:} Subregion of the observed 557\,GHz \water\ map of \roa. The scales of \vlsr\ and \ta\ are shown in the upper right corner. The {\it Herschel} beam at FWHM is indicated in the lower right corner (38\asec, grey). The red circle shows the size of the Bonnor-Ebert sphere model of SM\,1N discussed in the text. {\bf Lower:} Theoretical line profiles (red) are compared with the observed profiles (histograms). The spectra are half-beam spaced. The intensity is given in the \tmb\ scale, where $\eta_{\rm mb}=0.62$. The absorption dip is centred on \vlsr\,=\,+3.3\,\kms.}
  \label{obs_model}
\end{figure*}

A subregion of our \water\ 557\,GHz spectrum map is shown in Fig.\,\ref{obs_model}, where the protostar SM\,1N is shown by the red circle. Apparently, the basic feature of the characteristic line profile extends over a much larger region than that. A radial cut at half-beam spacing (about Nyquist sampled) is shown below the map, from which it is clear that the BE sphere model reproduces the observed line reasonably well at the infall centre. However, as one goes farther away from the centre, observation and model diverge increasingly (see the lower part of Fig.\,10). Evidently, a one solar mass collapsing protostar  alone cannot account for the observed widespread inverse P\,Cygni profiles.  It should also be clear that the observation of the spectrum towards merely one single position will not be sufficient to provide unambiguous evidence of protostellar mass infall.

\begin{figure}
 \resizebox{\hsize}{!}{
  \rotatebox{00}{\includegraphics{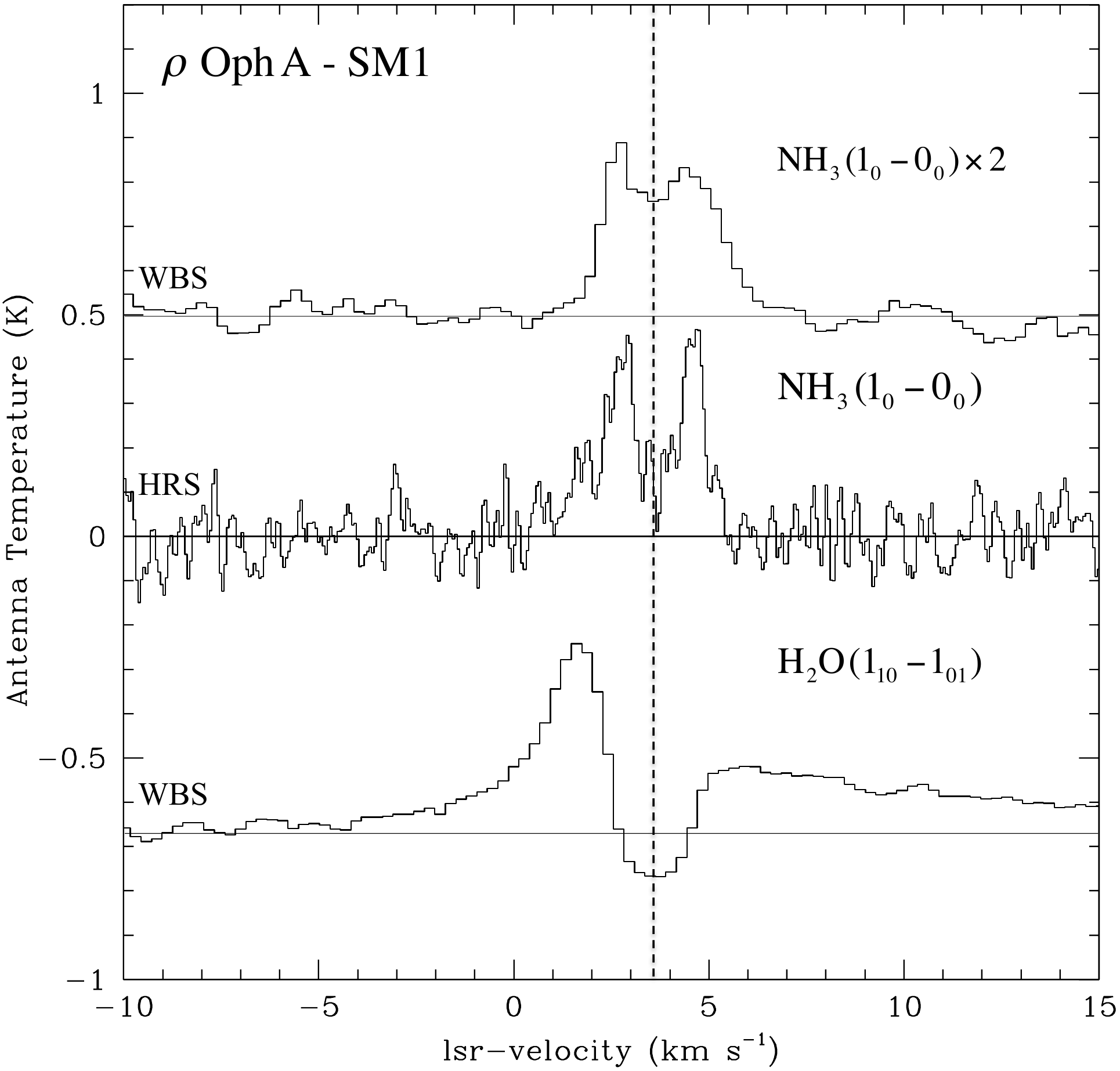}}
  }
  \caption{Ground-state lines of \ammonia\,572\,GHz and \water\,557\,GHz  observed at the same time in the different sidebands of HIFI. Both lines show profiles characteristic of infall, but the \ammonia\ line lacks the high-velocity wings seen in water due to the VLA\,1623 outflow. The middle figure shows the High Resolution Spectrometer (HRS) ammonia data, revealing the depth of the central absorption, and the other two the corresponding Wide Band Spectrometer (WBS) spectra at lower resolution.
        }
  \label{nh3_profile}
\end{figure}

\subsection{Theoretical models of infall: \ammonia}

\subsubsection{Line profiles: \ettnoll}

In contrast to the \water\ lines, the observed line profiles of \ettnoll\ are free from contaminating emission from the high-velocity gas in the outflow (Fig.\,\ref{nh3_profile}). The ammonia ground-state line can therefore be assumed to be potentially better suited to tracing protostellar infall.

\begin{figure*}
 \resizebox{\hsize}{!}{
  \rotatebox{00}{\includegraphics{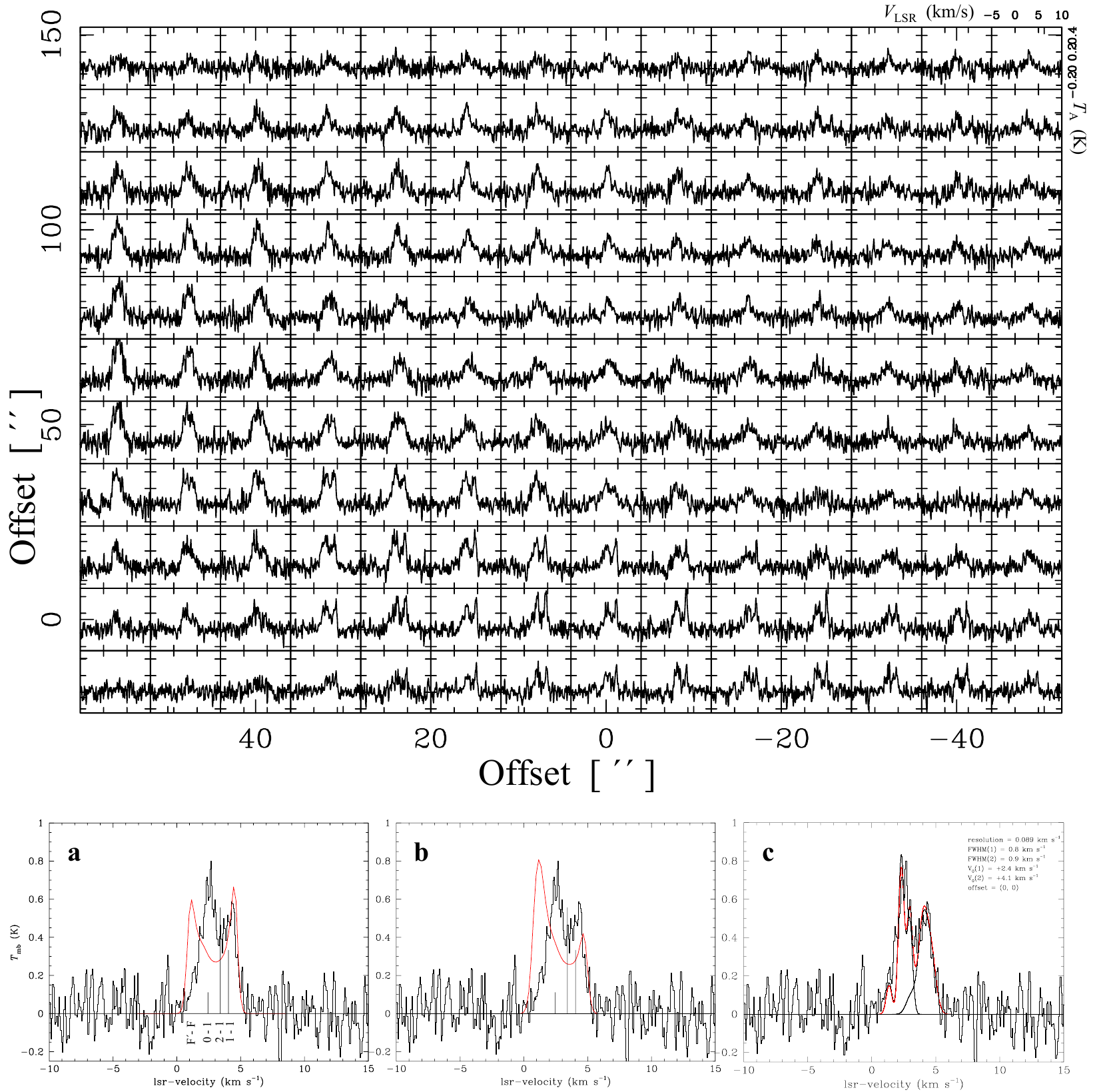}}
  }
  \caption{{\bf Upper:} Part of the observed {\it Herschel}-HIFI map in the ground state line of  o-\ammonia;  the scales are indicated in the upper right corner. {\bf Lower:}  Observed high-resolution (HRS) \ettnoll\ line profiles towards SM\,1N  shown  as histograms. The red lines show a static BE model in {\bf a}, whereas in {\bf b} an infall model is shown. The positions of the hfs components are shown as vertical bars, the lengths of which are normalized to 1.0, and their quantum numbers are indicated in {\bf a} (see Appendix\,B). {\bf c} shows the results of Gaussian hfs profile fitting for two velocity components, the parameters of which are inscribed.
        }
  \label{nh3_profiles}
\end{figure*}

In Fig.\,\ref{nh3_profiles}\,b, an infall model whose physical parameters are similar to  those for water has been computed for the \ettnoll\ line. The radiative transfer takes the overlap of the hyperfine structure (hfs) components explicitly into account. The highest optical depth of 240 is found at the frequency of the $F^{\prime} - F = 2 -1$ transition (see Appendix\,B), consistent with the observed absorption feature in the spectrum. Next to $F^{\prime} - F = 2 -1$ is the $F^{\prime} - F = 1 -1$  line that fits the observed red wing of the line. However, the theoretical emission peak is too low. This is in contrast to the blue peak that is well fitted in intensity, but there the wing is overestimated by the model. This appears to be a persistent feature of the computed profiles. Models that fit both the observed two-peak shape and the intensities have necessarily very high optical depths and the lines become broader. 

We have also tried to fit a static BE configuration (Fig.\,\ref{nh3_profiles}\,a).  As for the infall case, the ortho-\ammonia\ abundance is $1.5 \times 10^{-8}$, i.e. a factor of 35 higher than that found on the basis of our Odin observations with a 2\amin\ beam \citep{liseau2003}. Both the red wing and peak are well fit with this $\tau_0=260$ model; however, the blue peak intensity is much too low, and at the same time the blue wing is overestimated. The theoretical profile is too broad on the blue side, even though the turbulent speed is merely 0.3\,\kms. 

As for the \water\ lines, little compelling evidence for gravitational infall is provided by the ammonia line observations. Essentially perfect fits can be obtained by introducing two velocity components on an  ad hoc basis (Fig.\,\ref{nh3_profiles}c). Good examples would be hypothetical components at \vlsr\,=+2.4 and +4.1\,\kms, and with FWHM = 0.8\,\kms\ and 0.9\,\kms, respectively. However, on the basis of the analysis of \ntwohp\ spectral line maps \citet{liseau2015} found no evidence for different velocity components greater than 0.2\,\kms. We therefore abandon the results from Gaussian profile fitting as unreliable indicators of the velocity fields in the source.

\section{Conclusions}

In the densest and coldest regions of \roa, the abundance ratio of \molo\ to \water\ is of the order of ten. According to theoretical models of grain chemistry, $X$(\molo)/$X$(\water)$\,>1$ occurs only during brief periods of time, i.e. during intense \molo\ production in dense cores. This would limit the (chemical) age of SM\,1 to less than 30\,000 years and explain the elusiveness of \molo\ outside the solar system.

Although the line shapes of the ground state line of \etttio\ in \roa\ are textbook examples of protostellar infall, detailed modelling and radiative transfer calculations of the observed spatially extended emission makes this option unlikely. A similar conclusion is reached on the basis of accompanying \ettnoll\ observations.

Mapping the \ro\ cores A, B1, B2, C, D, E, and F in the \etttio\ line with Odin resulted in clear detections only in \roa. Upper limits are within 10 to 50\,mK (rms). 


\section{Epilogue: star formation in \roa}

\roa\ is a region of active low-mass star formation. It is special in some respects, but also shares  common properties in others. Based on a large observational material, a few conclusions that should also be valid  in a wider and general context can be drawn. Below we  summarise our findings and  combine the results of Papers I and II. 

\subsection{Age of the cloud}

Based on dynamical considerations, dense and cold molecular cores are generally believed to be very young. However, a proper calibration of these  qualitative age assessments is not readily achievable.

The chronometry of molecular clouds is a difficult task in general, but it may be achievable under special conditions. The hope is for `chemical clocks' that measure brief periods of transient times during the chemical evolution of the cloud. 

In the dense core SM\,1 in \roa, the highly unusual abundance ratio of \molo\ to \water, i.e. $X$(\molo)/$X$(\water)$\,>1$, indicates that the core is very young indeed; the age ranges from 5 000  to  30000 years. Outside this narrow window, abundance ratios equal to or larger than unity are rarely ever encountered in the ISM and rapidly approach observed upper limiting values.

However, given the association of young stellar objects with the cloud would suggest that \roa\ has produced stars for at least one million years, unless the stellar sources that happen to move in front of the cloud are interlopers (at a rate of 1\,pc per \kms) from other regions of the \roc. Proper motion data for the surroundings of \roa\ could help to settle this issue. 

For the more distant F core (\about\,0.8\,pc, see Fig.\,\ref{odin_map}), and partially also for E, \citet{ducourant2017} determined a mean proper motion that would correspond to an average lateral speed of 8\,\kms. If also representative  for the \roa\ neighbourhood, young objects (Class\,II to III) could have spilled in on time scales of less than \powten{5}\,yr, providing an age estimate for the cloud core of that order.

\subsection{Mass of the cloud} 

\subsubsection{Dust opacity}

It seems that it has generally been accepted that cloud masses are best determined from far-IR/submillimetre continuum observations, where the emission is likely optically thin and falls onto the Rayleigh-Jeans tail. Thus, to the first order, the only parameter to be determined appears to be the temperature of the dust. 

In reality, however,  the dust itself complicates things. In particular, the chemical composition and the size distribution of the grains determine the dust mass absorption coefficient $\kappa_{\nu}$,  which, as indicated by the subscript, is frequency dependent. This potentially distorts the spectrum of the cloud from being a pure black body. In Paper\,I, we examine this parameter in detail to determine its impact on the mass estimates. The most frequently exploited $\kappa_{\nu}$ values in the literature differ by factors of more than five.

\subsubsection{Gas-to-dust mass ratio}

To transform the estimated dust mass into total cloud mass requires the application of a gas-dust relationship, and an $m_{\rm gas}/m_{\rm dust}=100$ is commonly assumed, often  already implicitly in the value of $\kappa$. 

However, in cold cores, a constant value of one hundred will most likely not be true everywhere since much of the gas-phase molecules will locally be frozen onto the dust grains, decreasing this ratio.  Thus, potential freeze-out will invalidate commonly assumed calibrations,  e.g. $X$(CO)/$X$(\molh). In addition, gas and dust tracers will not measure the same parcels of material. For instance, in the cold core SM\,1 in \roa, the gas-to-dust mass ratio is down by one order of magnitude from its canonical value of one hundred, i.e. $(m_{\rm gas}/m_{\rm dust})_{\rm SM\,1}\,\sim 10$;  in addition, the projected distributions of the gas and dust do not coincide.

\subsection{Star formation efficiency}

Counting the stars that are associated with \roa\ is, in principle, a way to determine a limit to the cloud mass already converted into stars. Within a radius of \amindot{2}{0} (0.07\,pc) around VLA\;1623\,A, the SIMBAD database lists 30 young stellar objects (4 TT + 26 classified as YSO). Only a few have an assigned spectral type, which means that proper mass assignments are basically impossible for the majority of objects. T\,Tauri stars have statistically about half a solar mass and assuming that to be the case for all objects probably provides an upper limit to the stellar mass ($M_{\rm TT}+M_{\rm YSO} \sim 15$\,\msun) in \roa. Similarly, assuming that the objects listed as YSOs have a mass of 0.1\,\msun\ yields a strict lower limit, so  the total stellar mass probably falls within the range $4.5 < M_{\rm stars}/M_{\odot} < 15$. 

The upper limit is roughly of the same order of magnitude as estimates of the present \roa\ mass \citep[\about\,5 to 35\,\msun,][]{liseau2015}. The star formation efficiency, $M_{\rm stars} / (M_{\rm stars}+M_{\rm cloud})$, in \roa\ is likely less than 50\%; however,  if all  30 stellar objects have actually formed in \roa, it could be as high as  20\%. For the entire \roc\ complex (L\,1688), \citet{evans2009} determined an efficiency of 6\%, much less than the $\ge 22$\% of \citet{wilking1989}, but in line with the value estimated by \citet[][5-7\%]{liseau1995}.

\subsection{Evolution of the dust}

 \begin{figure*}
 \resizebox{\hsize}{!}{
  \rotatebox{00}{\includegraphics{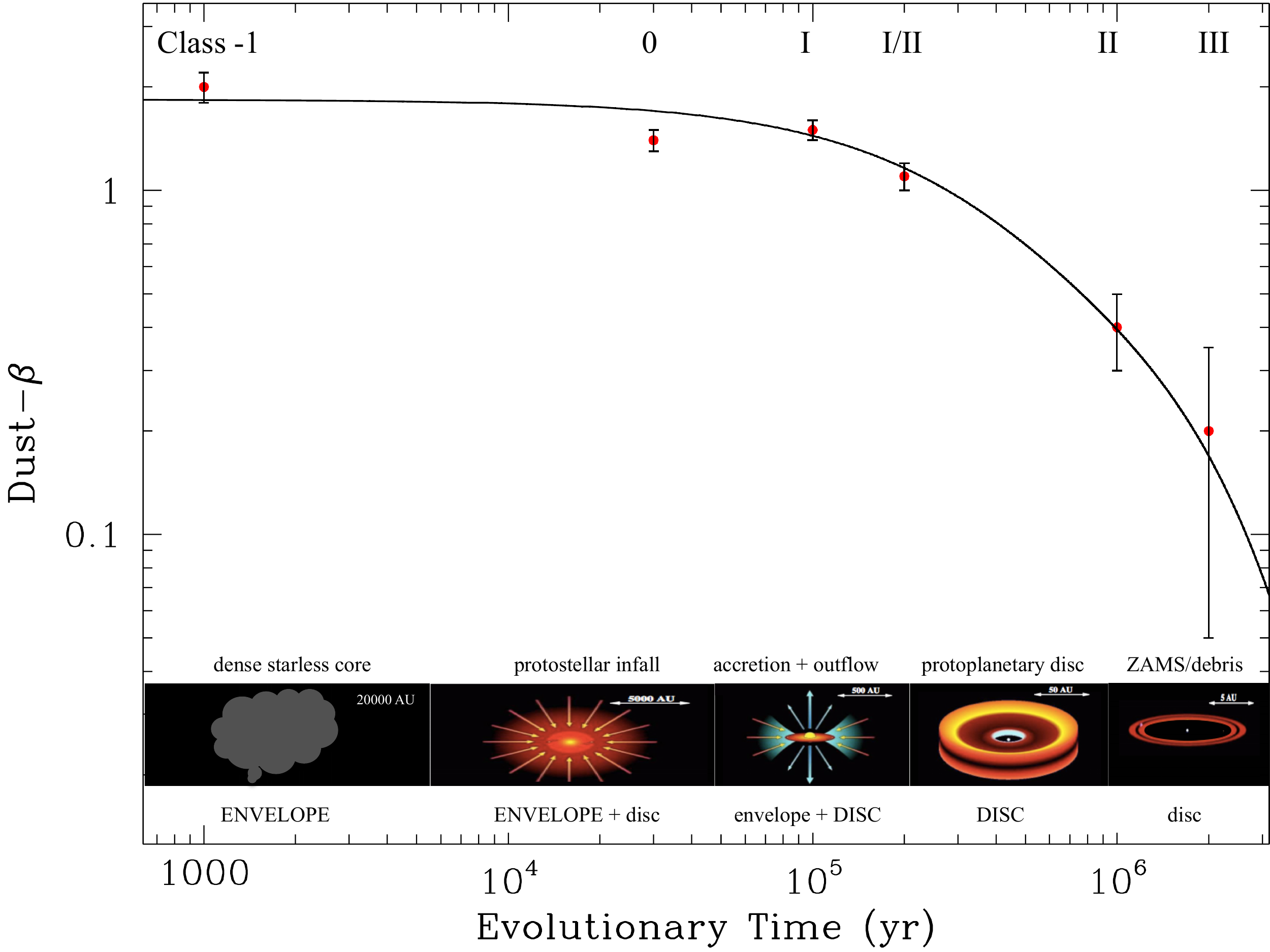}}
  }
  \caption{Dust opacity index $\beta$ as a function of evolutionary time of star formation. $\beta$ describes the frequency dependence of the dust opacity $\kappa_{\nu,\,{\rm dust}}$ in the Rayleigh-Jeans regime according to the power law $\kappa_{\nu,\,{\rm dust}} \propto \nu^{\,\,\beta}$. The cartoons are from McCoughrean/NASA/JWST. The major contributor to the dust emission at the various stages is indicated below the icons and the nature of the objects above them. The importance of a given contributor to the far-IR/submm emission is expressed in upper  and lower case letters, respectively. The common nomenclature of stellar formation, labelled Class $-1$ through Class\,III,  is shown above the data points with their $1 \sigma$ errors. The line through the data points is an analytical fit  of the form $\beta(t) = a_1\,\exp(-\gamma_1  t/\tau) + a_2\,\exp(-\gamma_2 t/\tau)$, where $\tau = 5\times 10^4$\,yr is a characteristic time scale, after which increased grain growth occurs in discs. 
        }
  \label{beta_t}
\end{figure*}

In \roa, there is a clear trend of  grain growth from the earliest to the latest phases of star formation (Fig.\,\ref{beta_t}):  big opacity exponents $\beta \sim 2$ mean small grains, whereas small $\beta \sim 0$ mean big grains. The dust grows from $a$\,\lapprox\,0.001\,\um\ during the initial dark cloud phase to $a$\,\gapprox\,100\,\um\ towards the later T\,Tauri and ZAMS phases. At intermediate times, i.e. during the protostellar and protoplanetary stages, grain sizes vary between 0.1 and 10\,\um. Respective grain size distribution parameters $p$, where $dn(a) \propto a^{-p}\,da$, are $4.5-5$ (Class$-1$ and 0),  $3.5 - 4$  (Class\,I and II), and 3\,(Class III). 

Further quantitative assessment is hampered by the poorly constrained early evolutionary time scales. However, if we assume currently adopted ages \citep[e.g.][]{evans2009}, the data can be fit by an exponential, as shown in Fig.\,\ref{beta_t}, from which we can infer that the observed dust is dominated by a population of very small grains at the earliest times (starless cores and dynamical collapse) up to  50 - 100 thousand years. After that, grains grow predominantly in circumstellar discs, as the relative contribution to the far-IR/submm emission from these grains continues to increase.  

\subsection{Protostellar mass infall}

Scientists with a talent for poetry once called the discovery of a collapsing protostar `the holy grail of star formation', and indeed, the identification of a true collapsing protostar in \roa\ appeared initially very promising \citep{ashby2000}. Like these authors, we used the observed \water\,557\,GHz line to fit its profile with a protostellar infall model and radiative transfer calculations, disregarding the extended line wings which were attributed to the VLA\,1623 outflow by \citet{ashby2000}.  In addition, we also applied the same model to the observed \ettnoll\ lines. These are not contaminated with outflow emission in the line wings. 

These models did indeed recover the infall signature imprinted in the \water\ line profile. However, our higher quality data and more advanced theoretical modelling could not uniquely confirm the proposed protostellar scenario as the observations show the infall profile over a much larger region than  the model with the slightly larger than 1\,\msun\ potential could account for.

\subsection{ PDR dynamics}

\roa\ is sandwiched  between the far-UV radiation fields from two B-type stars, one to the east and the other to the far west. The PDR parameter $G_0/n$ ranges from  0.001 to 0.01, considerably lower than for commonly studied PDRs. 
The spherical shell morphology of the eastern region could suggest that the dark core has been shaped and compressed by the radiation fields. However, currently we do not find convincing observational evidence for a strong pressure gradient at the PDR-border. Rather, it seems that an equilibrium situation has been achieved that has settled the gas into a quasi-static, clumpy configuration. 


\begin{acknowledgement} 
We thank the Swedish National Space Board (SNSB)  for its continued support of  our {\it Herschel} projects.  The contributions to this project in its initial phase by Drs.\,Per Bergman, Alexis Brandeker, and Michael Olberg are acknowledged. Dr.\,Tim-Oliver Husser and Dr.\,Emanuele Bertone are thanked for providing high-resolution spectra of stellar model atmospheres.
\end{acknowledgement}

\bibliographystyle{aa}
\bibliography{refs}

%
%
\appendix

\section{Odin observations of the \roc} 

Prior to {\it Herschel}, we had mapped the \roc\ in the \water\,($1_{10}-1_{01}$) line at 557\,GHz with the spectrometers aboard the Odin spacecraft \citep{frisk2003,nordh2003,olberg2003}.  Between 2003 and 2006, the \roc\ was mapped in the \water\,557\,GHz line during several observing runs (see Table\,\ref{obsrounds}). The coordinates of the map centre, i.e. at offset position (0,\,0), were R.A.\,=\,\radot{16}{26}{24}{6} and Dec.\,=\,\decdms{$-24$}{23}{54} (J2000). At the frequency of the \water\,($1_{10}-1_{01}$) transition, the beam width of Odin is 2\amin\ (HPBW) and for \roa, in particular, the spatial sampling was at about the Nyquist frequency, with relative offsets in 60\asec\ steps.

\begin{figure*}
  \resizebox{\hsize}{!}{
  \rotatebox{00}{\includegraphics{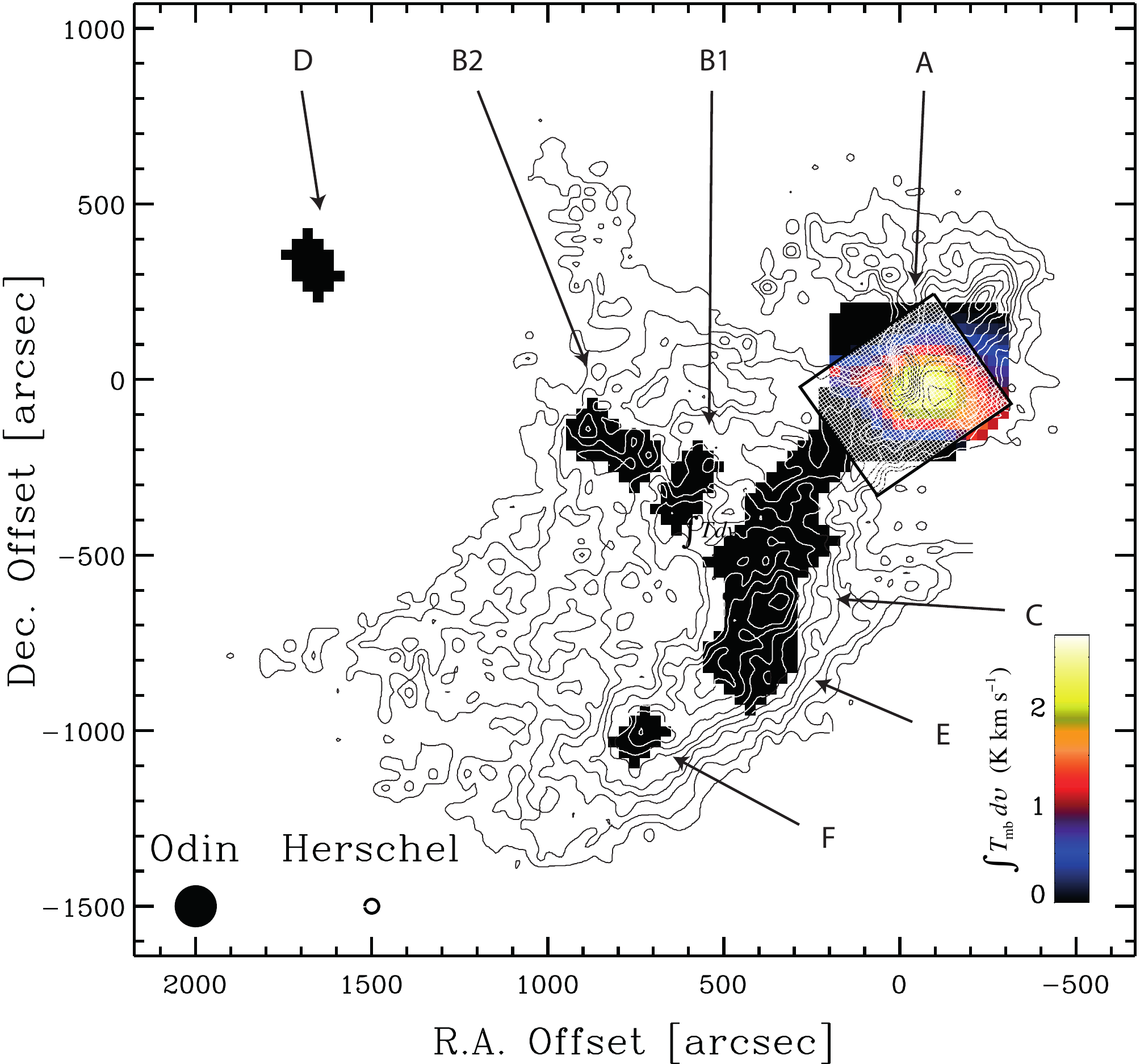}}
  }
  \caption{Spectral line image of the \roc\ (L\,1688) in the \water\,($1_{10}-1_{01}$) 557\,GHz line  obtained by Odin with its 2\amin\ beam, which is indicated to scale in the lower left corner. The colour-coding for the integrated intensity $\int \!\! T_{\rm mb}\,d\upsilon$ is shown by the scale bar in the lower right. The cores A through F are labelled, and the Odin data are superposed onto the contours of integrated $\rm C^{18}\rm O$ $(J=1-0)$ emission from \citet{umemoto2002}. The J2000-coordinates of the (0, 0) position are R.A.\,=\,\radot{16}{26}{24}{6} and Dec.\,=\,\decdms{$-24$}{23}{54} and offsets are in seconds of arc. The {\it Herschel} beam at 557\,GHz is shown next to that of Odin and the inclined, semi-transparent rectangle outlines the region mapped with HIFI.
        }
  \label{odin_map}
\end{figure*}

\begin{figure}
  \resizebox{\hsize}{!}{
  \rotatebox{00}{\includegraphics{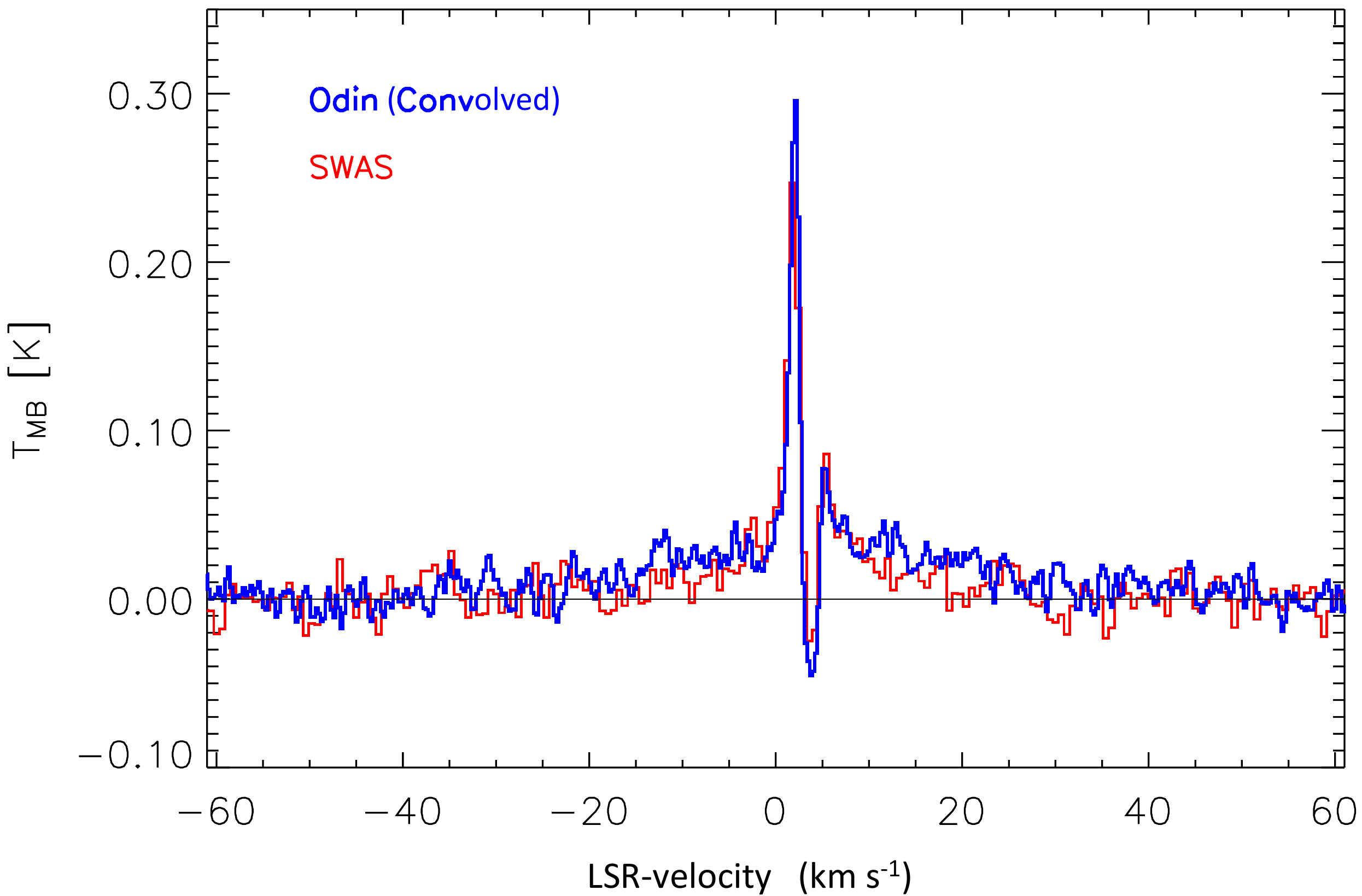}}
  }
  \caption{\water\,557\,GHz spectra of \roa\ obtained by SWAS and Odin. For this comparison, the Odin data have been convolved into the larger SWAS beam and are shown in blue, whereas the SWAS spectrum at two times lower frequency resolution is displayed in red. The comparison reveals a high degree of consistency and shows the high-velocity wings, traceable to at least $\pm 30$\,\kms\ relative to the systemic Doppler velocity.
        }
  \label{odin_swas}
\end{figure}

\begin{table*}[ht]
  \caption[]{Journal of Odin \water\,557\,GHz observations}
  \label{obsrounds}
  \centering          
  \begin{tabular}{l l c c c c l}     
  \hline\hline        
       \noalign{\smallskip}
       Year  &  Date                                         & Orbits           & Used Orbits     &        Integration (hr) &  Pointings          & Source \\
       \noalign{\smallskip}
       \hline
       \noalign{\smallskip}
       2003  &  February 1 - 24                       & 10577 - 10918  &         120     &         36.2          &          12           & $\rho$\,Oph A\\
       2003  &  August 8 - 18                 & 13372 - 13527  &\phantom{1}83&         20.9            &\phantom{1}4           & $\rho$\,Oph A\\
       2004  &  January 31                             & 15992 - 16006  &\phantom{1}11&\phantom{1}3.3         &\phantom{1}2           & $\rho$\,Oph A\\
       2004  &  August 1 - 12                 & 18730 - 18901   &\phantom{1}86&         18.9            &        3 + 5          & $\rho$\,Oph B1, B2 \\
       2005  &  July 31 - October 1           & 24181 - 25100   &          394   & 73.5 + 109.4$^a$      &          51           & $\rho$\,Oph A to F \\
       2006  &  February 2 -  March 2         & 26956 - 27370   &       198     &         75.4            &          12           & $\rho$\,Oph A\\
       2006  & August 26 - September 18  & 30016 - 30357   &               142   &  48.3 + 46.6$^b$      &\phantom{1}3           & VLA 16293  \\
       \hline
       \hline
        \noalign{\smallskip}
       Total &                                               &                               &    1034   &        432.5          &        92                     &                          \\
       \noalign{\smallskip}
       \hline
\end{tabular}
\tablefoot{    
$^{(a)}$ OB + OC;  $^{(b)}$ VLA Outflow: AOS + AC
}
\end{table*}

The front- and backends \citep{frisk2003,olberg2003} were, respectively, the 555\,B1 receiver and either of the two digital autocorrelators (ACs) with a bandwidth of 400\,MHz. The channel resolution was 0.5\,MHz, corresponding to 0.27\,\kms\ and implying a velocity resolution of \about\,0.5\,\kms. However, for the large map of 2005 and  the outflow strip scan of 2006, two \water\ receivers were used and the backends were the other autocorrelator (AC2 in 2005) and the acousto-optical spectrometer (AOS in 2006). The observing mode was always Dicke-switching, with the \adegdot{4}{4} sky-beam pointing off by $42$\adeg\ \citep{olberg2003}.

During the mapping procedure in 2005, all 51 positions were observed once per satellite revolution in a raster scan mode. Individual spectral scans were 5\,s each. An `off-position', supposedly free from molecular line emission, was also observed once per revolution. The total on-source time spent on every observed point was on  average 3.5 hours for the cores C to F. For the B cores, the average observing time was 6 hours on-source. Finally, for \roa, this time was about 12 hours. 

The system noise temperature \tsys\  was within the range 3\,300\,K to 3\,400\,K (single sideband, SSB). Examination of all the position data reveals that the relative pointing accuracy
is better than 2\asec\ (rms scatter) for any individual position and better than 5\asec\ for the overall final map ($<4\%$ of the Odin beam width).

The basic data reduction procedure is described by \cite{olberg2003}. The appropriate equations are given in that paper and will not be repeated here. The spectrum of the off-position, obtained once per revolution, was used to monitor the reduction of the Dicke-switched spectra along the pipeline. This procedure assured that the final results are reliable. The absolute accuracy of the intensity scale is believed to be better than 10\% \citep{hjalmarson2003}. Comparison with the data obtained by SWAS shows excellent agreement (Fig.\,\ref{odin_swas}).

Due to the relatively low orbit of Odin, astronomical sources are generally occulted by the Earth. As a consequence, Odin `is observing' through the Earth's atmosphere twice per orbital revolution, i.e. during ingress  and egress phases of the eclipse \citep[see Fig.\,2 in][]{nordh2003}. This circumstance has  an advantage, as the telluric ground state line of \water\ at 556.936\,GHz permits a highly accurate frequency calibration \citep{larsson2003}. In fact, the final accuracy is better than half a channel width ($\delta \upsilon < 0.14$\,\kms), which takes into account the {averaging of all} the scans for the individual map positions, corrected for the motion of the Earth and the satellite in the local standard of rest (LSR) velocity scale.

With Odin, the average of the scans for each position results in a 70-point map for ortho-water with 60\asec\ spacing of the dense cores A to F. As can be seen in Fig.\,\ref{odin_map}, no water emission was detected towards any of the cores except towards \roa. Furthermore, Table \ref{avespectab} shows that the overall achieved sensitivity is about 40\,mK per pointing\footnote{The conversion from K to Jy is about 0.02\,mK/Jy \citep{sandqvist2003}.}. In individual cases, e.g. core C, an rms noise temperature of \trms\,=\,10\,mK is obtained when averaging the measurements for all positions towards a single core.  In Table\,\ref{avespectab}, the basic results are summarized. Of the seven cores observed, only \roa\ was clearly detected (Fig.\,\ref{odin_map}). The ground-state transition of ortho-water in \roa\ has also been observed with SWAS \citep{ashby2000,snell2000}. The Odin and SWAS data are in excellent agreement (Fig.\,\ref{odin_swas}).

\begin{table}[ht]
  \caption[]{Odin \water\, 557\,GHz results for \ro\ cores A through F}
  \label{avespectab}
  \centering          
  \begin{tabular}{lcccc}     
  \hline\hline        
       \noalign{\smallskip}     
       \ro\       &  Pointings          & Integration   & $T_{\rm rms}^{\rm point}$& $T_{\rm rms}^{\rm ave}$ \\
        Core    &                               & (hour)        & (mK)           & (mK)       \\
       \noalign{\smallskip}
       \hline
       \noalign{\smallskip}
       A        &   35                                & 155       & 20                  & \phantom{1}6  \\
       B1       & \phantom{1}3          & 3.6   & 49            & 49  \\
       B2       & \phantom{1} $\!5$             & 31.0          & 39            & 18  \\
       C        & 16                                    & 64.0          & 39              & 10  \\
       D        &  \phantom{1}2                 & 6.4           & 41            & 29  \\
       E        &  \phantom{1}9                 & 36.0          & 40            & 13  \\
       F        &  \phantom{1}2                 & 3.0   & 66            & 47  \\
       \noalign{\smallskip}
       \hline
\end{tabular}
\end{table}


\section{Hyperfine structure in \ammonia}

\begin{figure*}
  \resizebox{\hsize}{!}{
  \rotatebox{00}{\includegraphics{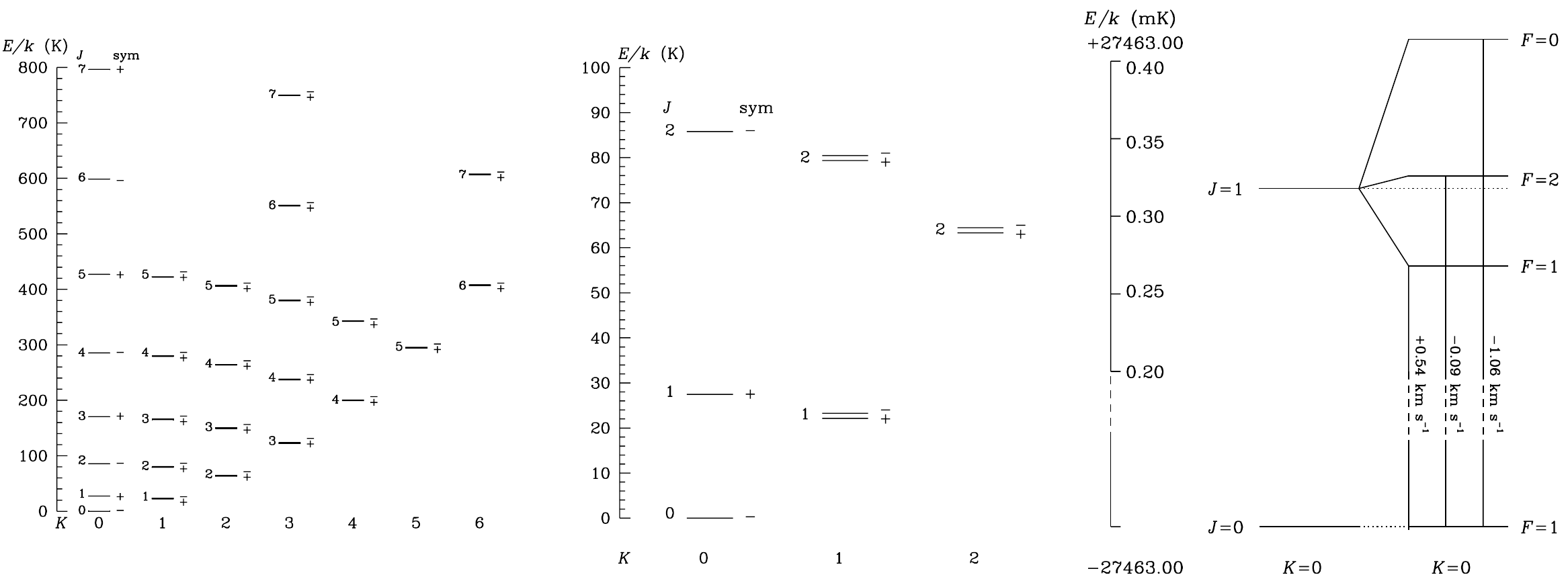}}
  }
  \caption{{\bf Left:} Rotational energy levels of \ammonia\ \citep[see the LAMDA data bank,][]{schoier2005}. Energies up to 800\,K are given for the ortho-\ammonia\ states ($K=0$ or $K=$ multiple of 3), whereas these are roughly half for para-\ammonia\ ($K \ne 0$ and $K\ne$ multiple of 3). All levels are plotted to scale, rendering the inversion splittings unresolved. {\bf Middle:} Same as left frame, but for energies below 100\,K. For symmetry reasons, half of the inversion levels do not exist for $K=0$. In this panel, the inversion levels, shown to scale, are resolved. {\bf Right:}  Hyperfine splitting, due to quadrupole interaction, of the $(J=1,\,K=0)$ level is shown. The resulting shifts, in \kms, for the \ettnoll\ line are indicated next to the components. The drawing is to scale and energies are in milli-Kelvin. 
  }
  \label{E_lev}
\end{figure*}

The rotational levels of \ammonia\ are hyperfine-split due to quadrupole interaction (see Fig.\,\ref{E_lev}).  The relative line strengths in equilibrium are 

\begin{eqnarray}
S_{\rm ul}^{\rm hfs} = \left [ (2 J_{\rm u}+1)(2 J_{\rm l}+1)  \left (  \begin{array}{lcr} J_{\rm l}\; \;\;1 \;\;\;J_{\rm u} \\
                                                                                                                        K \; \;  \;        0\;  -K   \end{array} \right )^2 \right ]     \times    \nonumber \\                                                                                                                                                                                  
 \left [ (2 F_{\rm l} +1)(2 F_{\rm u} +1)  \left \{  \begin{array}{lcr} J_{\rm u}\; \;\;    F_{\rm u} \;\;\;I \\
                                                                                                                                          F_{\rm l}\; \;  \;   J_{\rm l}\;\;\; \;1  \end{array} \right \}^2 \right ] ,                                                                                                                                      
\end{eqnarray}

where the first factor in square brackets is the line strength $S_{\rm ul}$ for rotational transitions ($J_{\rm u},\,K \rightarrow J_{\rm l},\,K$). The relative strengths for the hyperfine transitions of the ($1,\,0 \rightarrow 0,\,0$) line are given in Table\,\ref{hfs_data}. 

Ammonia comes in two `flavors', namely ortho-\ammonia\ (or A-state)  and para-\ammonia\ (or E-state). The statistical weights for the rotational levels are $g_{_J}\,g_{_I}$, where $g_{_J}=2J+1$ and where $g_{_I}$ refers to the spin of the three hydrogen nuclei, $I=I_{\rm H}$. These are \citep{townes1955} \\
\\
A-state or ortho-\ammonia: $I=\frac {3}{2}; \,g_{_I} = \frac {1}{3} (2I + 1)(2I - 1)I = 4$ \\
E-state or para-\ammonia: $I=\frac {1}{2};\,g_{_I} = \frac {1}{3} (2I + 1)(4I^2 + 4I) = 2$\\
\\
so that the equilibrium ratio of the weights of the ortho-to-para states is $o/p=2$. At high temperatures, $o/p=1$ \citep{faure2013}. For the ortho-states, $K$ takes the values $K=3n$, $n=0,1,2,\dots,N$ and for the para-states all the others. Therefore, the results for the $(1_0-0_0)$ ortho-line reported here are not necessarily directly related to those for the inversion lines, e.g. (1, 1), (2, 2), etc., that have been commonly cited in the literature. 

The statistical weights of the hfs lines are $g_{\rm F}=2 F + 1$, and the Einstein $A$-values are obtained from 
                                                       
\begin{equation}
A(J, J-1)_{K, \pm} = \frac {64\,\pi^4 \nu^3 \mu^2}{3\,h c^3 } \,\,\frac {J^2-K^2}{J(2J+1)}
\end{equation}

For NH$_3$, the dipole moment $\mu = 1.476 \pm 0.002$\,Debye \citep{poynter1975}. The hfs levels and their weighted Einstein $A$s have been implemented in the ALI programme, which includes 49 levels, up to $\sim 600$\,K, and 104 radiative transitions. When $\tau \ge 1$ at line centre, the relative contributions of the hfs lines will be different from those in equilibrium. Line overlap is explicitly accounted for by the transfer code.

\begin{figure}
  \resizebox{\hsize}{!}{
  \rotatebox{00}{\includegraphics{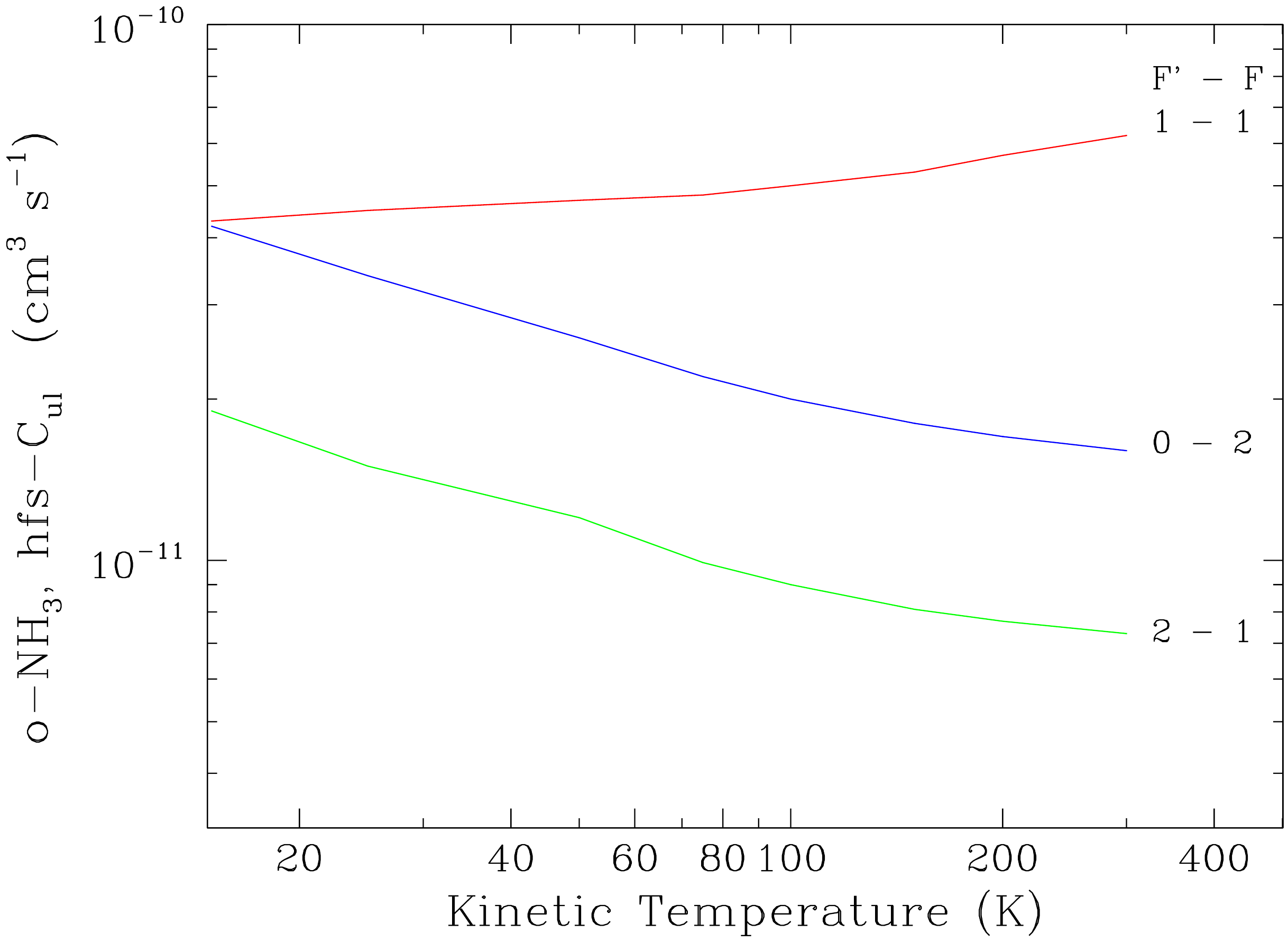}}
  }
  \caption{Rate coefficients for collisional de-excitation $C_{\rm ul}$ for hfs transitions of $J,\,K = 1,\,0 - 0,\,0$ (see  text).
  }
  \label{nh3_coll}
\end{figure}

Collision rate constants $C_{\rm u \leftrightarrow l}$ for the hfs transitions $J_{\rm u \leftrightarrow l} =1 \leftrightarrow 0$, $K=0$ have been published by \citet{chen1998}. However, for the sake of consistency, we computed the collision rates for the hfs transition among all {\it J},  {\it K} levels according to the prescription given by \citet{alexander1985}. These approximations are valid within a given $K$-ladder (see e.g. Fig.\,\ref{nh3_coll}). For cross-$K$ transitions we simply adopted the data provided by \citet{danby1988}. Included are 1162  transitions for eight temperatures in the range 15 to 300\,K.

\begin{table}
\begin{flushleft}
 \caption{\label{hfs_data} Data for \ammonia\ quadrupole interaction for $K=0,\,J=0$.}
\begin{tabular}{llcccl}
\hline\hline
\noalign{\smallskip}
hfs                             & Frequency             & $\Delta \upsilon_{F^{\prime}}$         & $(E_0 - E_{F^{\prime}})/k$    &Relative                \\ 
$F^{\prime} - F$        & (MHz)                      & (\kms)                                                   &    (K)                                          & Strength       \\
\noalign{\smallskip}    
\hline
\noalign{\smallskip}    
$1 -1$                  & 572\,497.1247         & \phantom{1}$+0.542$                   &$+ 5.0 \times 10^{-5}$             & 0.333    \\
$2 -1$                  & 572\,498.3391 & \phantom{1}$-0.094$                   &$- 8.1 \times 10^{-6}$             & 0.556     \\
$0 -1$                   & 572\,500.1914        & \phantom{1}$-1.064$                   & $- 9.7 \times 10^{-5}$          & 0.111     \\  
\noalign{\smallskip}
\hline
  \noalign{\smallskip}
\noalign{\smallskip}    
\end{tabular}
Adopted centre frequency is $\nu_0 = 572\,498.160$\,MHz and $eqQ = - 4.08983$\,MHz \citep{cazzoli2009}.  
\end{flushleft}
\end{table}

\section{CS line maps with the Swedish ESO Submillimetre Telescope }

Mapping observations in CS\,(2-1), (3-2), and (5-4) were performed at the 15\,m Swedish ESO Submillimetre Telescope \citep[SEST,][]{booth1989} during 1996 and 1997 (Table\,\ref{SEST}). As frontends we used SIS-mixer receivers and as backend a 1000-channel acousto-optical spectrometer (AOS) with 43\,kHz resolution. The observations were performed in frequency switching mode with a frequency throw of 7\,MHz for the 100 and 150\,GHz observations and of 15\,MHz for those at 250\,GHz. At these frequencies system noise temperatures were \tsys\,\about\,150, 200, and 400\,K, respectively. 

The pointing of the telescope was regularly checked using stellar SiO masers and found to be better than 3\asec\ rms. The (2-1) and (3-2) maps were obtained simultaneously with two receivers and these covered an area of 5\amin\,$\times$\,5\amin. In the (5-4) line, a $\Delta \alpha \times \Delta \delta = 3^{\prime} \times 2^{\prime}$ was obtained. The data are internally chopper-wheel calibrated in the \tas-scale \citep{ulich1976} and as celestial calibrator,  we used M\,17\,SW (R.A. =18h 20m 23.1s, Dec. = $-16$\adeg\ 11\amin\ 43\asec, J2000.0).  Repeated observations on different occasions resulted in stable intensities of the lines within 4 to 5\%. The \tmb\ values were obtained by applying the main beam efficencies, $\eta_{\rm mb}$, provided in Table\,\ref{SEST}.

Standard data reduction techniques were applied in Class and with the locally available software package \texttt{xs}\footnote{\texttt{ftp://yggdrasil.oso.chalmers.se/pub/xs/}}, involving folding of the spectra, fitting and subtracting base lines, and averaging multiple scans for the same position. 

\begin{table}[h]
\caption{Instrumental reference: SEST}             
\label{SEST}      
\begin{tabular}{lcccccc}    
\hline\hline    
\noalign{\smallskip}             
        CS              &   $\nu_0$                     & $E_{\rm up}/k$         & $A_{\rm ul}$          &  HPBW & $\eta_{\rm mb}$        \\
$J+1 - J$               &   (GHz)                               & (K)                           & (s$^{-1}$)              &  (\asec)      &                               \\
\noalign{\smallskip}    
\hline                        
\noalign{\smallskip}    
\noalign{\smallskip}    
$2-1$   &\phantom{1} 97.9809533& 7                      & $1.68 \times 10^{-5}$ &       50      & 0.73     \\
$3-2$   &  146.9690287  & 14                            &  $6.06 \times 10^{-5}$        &       33      & 0.67    \\      
$5-4$   &  244.9355565  & 35                            &  $2.98 \times 10^{-4}$        &       20      & 0.60    \\
\noalign{\smallskip}    
\hline                                   
\end{tabular}
\end{table}

\section{Synopsis of gas observations: lines and telescopes }

In Table\,\ref{obs_sum} we provide an overview of the spectral lines that have been observed and analysed in Papers I and II.

\begin{table}[h]
\caption{Summary of gas observations}             
\label{obs_sum}      
\begin{tabular}{llcrl}    
\hline\hline    
\noalign{\smallskip}             
Atom     or             & Quantum       &    Frequency          & Upper level         & Observing              \\
Molecule                & Numbers       &   (GHz)                       & Energy (K)      & Facility                      \\
\noalign{\smallskip}    
\hline                        
\noalign{\smallskip}    
\noalign{\smallskip}    
\molh                   &$J=2-0$, S(2)                                          & $244 \times 10^6$       &1682   & ISO-CAM CVF \\
                                &$J=3-1$, S(3)                                  & $310 \times 10^6$       &2503   & ISO-CAM CVF \\
                                &$J=5-3$, S(5)                                          & $433 \times 10^6$ &4586 & ISO-CAM CVF \\
O                               &$J=1-2$                                                        & 4744                    & 228   & {\it Herschel}-PACS \\
                                &$J=0-1$                                                        & 2010                    & 326   & {\it Herschel}-PACS \\
CH$^+$                  &$J=2-1$                                                        & 1669                    & 153   & {\it Herschel}-HIFI   \\                              
\ammonia                        &$J=1-0,\,K=0-0$                                        &  572                    &   27  & Odin,  {\it Herschel}-HIFI \\ 
\water                  &$J=1-1,\,K_{\rm a}=1-0,\,K_{\rm c}=0-1$        & 556                     &   37  & Odin,  {\it Herschel}-HIFI    \\
                                &$J=2-1,\,K_{\rm a}=1-0,\,K_{\rm c}=2-1$         & 1669                  & 114   &  {\it Herschel}-PACS, {\it Herschel}-HIFI     \\              
C$_2$H                  & $J=17/2-15/2$, $N=9-8$                        & 785                     & 208   & {\it Herschel}-HIFI   \\      
                                & $J=19/2-17/2$, $N=9-8$                        & 785                     & 208   & {\it Herschel}-HIFI   \\                              
$^{13}$C$^{16}$O        &$J=7-6$                                                        & 771                     & 148   & {\it Herschel}-HIFI   \\
$^{12}$C$^{17}$O        &$J=7-6$                                                        & 786                     & 151   &  {\it Herschel}-HIFI  \\      
\ntwohp                 &$J=3-2$                                                        & 279                     &   28  & APEX \\
                                &$J=6-5$                                                        & 558                     &   94  & {\it Herschel}-HIFI   \\
\molo                   &$J=1-1,\,N=1-0$                                        & 118                     &     6 & Odin \\
                                &$J=3-1,\,N=3-2$                                        & 487                     &   26  &  {\it Herschel}-HIFI  \\
                                &$J=5-3,\,N=4-4$                                        & 773                     &   61  & {\it Herschel}-HIFI   \\
$^{12}$C$^{32}$S        & $J=2-1$                                                       &   97                    &    7          &  SEST  \\
                                & $J=3-2$                                                       & 146                     &  14           &  SEST \\      
                                & $J=5-4$                                                       & 244                     &  35   &  SEST \\
                                &$J=10-9$                                               & 489                     &129            & {\it Herschel}-HIFI   \\
\noalign{\smallskip}    
\hline                                   
\end{tabular}
\end{table}

%
%
\end{document}